\documentclass[fleqn,usenatbib]{mnras}

\usepackage[T1]{fontenc}
\usepackage{ae,aecompl}

\usepackage{graphicx}	
\usepackage{amsmath}	
\usepackage{amssymb}	

\usepackage{natbib}
\bibpunct{(}{)}{;}{a}{}{,}

\newcommand{\be}{\begin{equation}}
\newcommand{\ee}{\end{equation}}
\newcommand{\bea}{\begin{eqnarray}}
\newcommand{\eea}{\end{eqnarray}}

\newcommand{\m}{\,\hbox{m}}
\newcommand{\mm}{\,\hbox{mm}}
\newcommand{\km}{\,\hbox{km}}
\newcommand{\mum}{\,\mu\hbox{m}}
\newcommand{\cm}{\,\hbox{cm}}

\newcommand{\AU}{\,\hbox{au}}
\newcommand{\g}{\,\hbox{g}}
\newcommand{\s}{\,\hbox{s}}

\newcommand{\yr}{\,\hbox{yr}}
\newcommand{\Myr}{\,\hbox{Myr}}
\newcommand{\Gyr}{\,\hbox{Gyr}}

\newcommand{\K}{\,\hbox{K}}

\title[Masses of debris discs]{Solution to the debris disc mass problem:\\
planetesimals are born small?}

\author[Krivov and Wyatt]{
Alexander V. Krivov$^{1}$\thanks{E-mail: krivov@astro.uni-jena.de (AVK)}
and Mark C. Wyatt$^{2}$
\\
$^{1}$Astrophysikalisches Institut und Universit\"atssternwarte, Friedrich-Schiller-Universit\"at Jena,
      Schillerg\"a{\ss}chen~2--3, 07745 Jena, Germany\\
$^{2}$Institute of Aastronomy, University of Cambridge, Madingley Road, Cambridge CB3 0HA, UK
}

\date{Accepted 2020 August 6. Received 2020 August 5; in original form 2020 June 10}

\pubyear{2020}

\begin{document}
\label{firstpage}
\pagerange{\pageref{firstpage}--\pageref{lastpage}}
\maketitle

\begin{abstract}
Debris belts on the periphery of planetary systems, encompassing the
region occupied by planetary orbits, are massive analogues of
the Solar system's Kuiper belt.
They are detected by thermal emission of dust released in collisions
amongst directly unobservable larger bodies that carry most of the debris disc mass.
We estimate the total mass of the discs by extrapolating up the mass of emitting dust 
with the help of collisional cascade models.
The resulting mass of bright debris discs appears to be
unrealistically large, exceeding the mass of solids available in the systems at the preceding
protoplanetary stage.
We discuss this ``mass problem'' in detail and investigate possible solutions to it.
These include uncertainties in the dust opacity and planetesimal strength,
variation of the bulk density with size,
steepening of the size distribution by damping processes,
the role of the unknown ``collisional age'' of the discs,
and dust production in recent giant impacts.
While we cannot rule out the possibility
that a combination of these might help, we argue that the easiest solution would be to assume that
planetesimals in systems with bright debris discs were ``born small'', with sizes in the kilometre range,
especially at large distances from the stars.
This conclusion would necessitate revisions to the existing planetesimal formation models, and may have
a range of implications for planet formation. We also discuss potential tests to constrain
the largest planetesimal sizes and debris disc masses.
\end{abstract}




\section{Introduction}

Planetary systems form in, and from, gaseous discs around young stars.
After the cooling phase, these protoplanetary discs contain about
one percent of their material in the form of solids that condensed out of gas.
The mass of a protoplanetary disc in general (and a fraction of mass in
solids) is one of the key parameters that largely determine
which planetesimals and then planets form in the system,
how they do it and when, 
and set the architecture of the emerging planetary system.
However, the mass of solid material in such discs, inferred from (sub)mm
observations, seems to be lower than the estimated mass of solid cores
of known extrasolar giant planets, which is known as the
``disc mass problem''
\citep{greaves-rice-2010,williams-2012,najita-kenyon-2014,mulders-et-al-2015,%
manara-et-al-2018,tychoniec-et-al-2020}.
Various ways of mitigating the problem have been proposed,
for instance that planets form earlier than commonly assumed
\citep[e.g.,][]{williams-2012,tychoniec-et-al-2020},
the dust mass in protoplanetary discs is highly underestimated
\citep[e.g.,][]{nixon-et-al-2018,zhu-et-al-2019},
or that these discs are continuously replenished from the environment
\citep[e.g.,][]{manara-et-al-2018}.
Yet a compelling solution is still pending.

After the gas dispersal at several Myr, a system leaves behind
a set of planets, as well as one or more dusty belts of leftover
planetesimals that, for some reasons, have failed to grow to
full-sized planets in certain radial zones \citep[e.g.,][]{wyatt-2008, matthews-et-al-2013, hughes-et-al-2018}. These are called debris discs.
Debris belts on the periphery of the planetary systems, encompassing the
region occupied by known or alleged planets, are massive analogues of
the Solar system's Kuiper belt. Like with the protoplanetary discs,
the mass of the debris discs in the systems is a key to understand
the evolution of debris discs and their interaction with planets.

Unfortunately, the total mass of a debris disc is difficult
to determine. The only quantity that can be derived from observations is
the \emph{dust} mass \citep[e.g.,][]{holland-et-al-2017}. The total mass of a debris disc must be dominated
by the unobservable dust parent bodies, planetesimals.
Thus the standard way to access the total disc mass is to invoke
theoretical models of the collisional cascade to extrapolate the dust
mass to the total mass of the disc.
This procedure uncovers a problem similar to that for the
protoplanetary discs: the inferred total masses of some debris discs appear
to be unrealistically large, by far exceeding the masses of solids in their
progenitors, protoplanetary discs \citep{krivov-et-al-2018}.

There is a related problem in the Solar system's asteroid belt.
While there are good constraints on its current mass,
because the largest planetesimals can be seen individually,
its initial mass can only be constrained by interpreting the current size distribution
within the context of collisional cascade models.
Such models conclude that asteroids were likely born big
\citep[100s of km in size,][]{bottke-et-al-2005,morbidelli-et-al-2009b,delbo-et-al-2017}.
Here we argue that, in contrast, the solution to the debris disc mass problem
is that planetesimals were born small.

This paper starts with  explaining the ``debris disc mass problem''
in detail (Sect.~\ref{s:problem}) and presents conceivable ways of resolving
the conundrum (Sect.~\ref{s:solutions}).
Section~\ref{s:discussion} contains a discussion,
and Sect.~\ref{s:conclusions}
summarizes our conclusions. 

\section{The problem}
\label{s:problem}

\subsection{The minimum mass of a debris disc}
\label{ss:min mass}

On timescales comparable to their ages, debris discs gradually lose their mass
\citep{dominik-decin-2003,wyatt-et-al-2007,loehne-et-al-2007}.
This happens because the collisional cascade grinds planetesimals all the way down
to dust particles, the smallest of which are either blown out of the systems by direct
radiation pressure or sublimate in the vicinity of the star,
being transported there by drag forces.
As a result, the initial mass of any debris disc should exceed its present mass by an amount equal to the total mass lost
over its age.

Consider a disc containing objects with sizes up to $s_\mathrm{max}$, which has
a total mass $M_0$ initially. Let $\tau_0$ be the collisional lifetime of the largest
objects of radius $s_\mathrm{max}$ at that initial epoch.
If the disc evolves in a quasi-steady state, and if the pairwise collisions are the only
mechanism that creates and removes the particles from the disc,
then the disc mass at a later time $t$ is given by
\citep{wyatt-et-al-2007,loehne-et-al-2007}
\be
M = M_0 / \left( 1 + t/\tau_0 \right) .
\label{eq:M(t)}
\ee
With this decay law, the initial disc mass can be calculated from 
the mass loss rate $\dot{M}$ at its current age $t$:
\be
M_0 = |\dot{M}| \;\tau_0 \; \left( 1 + t/\tau_0 \right)^2 .
\label{eq:M-Mdot}
\ee

It can usually be assumed that the collisional cascade in a debris disc is fed by planetesimals
of a size for which their collisional lifetime is equal to the age of the system
\citep[see discussion in Sect. 5.3 of][]{wyatt-dent-2002}.
Accordingly, we can choose the maximum size $s_\mathrm{max}$ in such a way that $\tau_0 = t$.
For bright debris discs around $\sim \Gyr$-old stars, this corresponds to
$s_\mathrm{max} \sim 1\km$, see Sect.~\ref{ss:recent ignition}.
Planetesimals of larger sizes may be present in the system and, if so, could dominate the total
mass of the disc.
Therefore, setting $\tau_0 = t$ gives us a {\em lower} limit on the
initial disc mass $M_0$.
In this case, Eq.~(\ref{eq:M(t)}) shows that the disc has lost half of its initial mass
by the time $t$, to wit, $M = M_0/2$, and Eq.~(\ref{eq:M-Mdot}) gives
\be
M_0 \geq 4 |\dot{M}| \;t .
\label{eq:M-Mdot simple}
\ee

The mass loss rate $|\dot{M}|$ can be inferred from the observables.
\citet{matra-et-al-2017b} showed that it 
can be computed from the disc fractional luminosity, radius, and width
(their Eq.~21), assuming an ideal collisional cascade \citep{dohnanyi-1969}.
Applying their formula to the Fomalhaut disc,
they derived the mass loss rate of $11 M_\oplus \Gyr^{-1}$.
Using Eq.~(\ref{eq:M-Mdot simple}) for the age of Fomalhaut of $440\Myr$ results in
$M_0 \geq 19 M_\oplus$ \citep[similar to the conclusion of][]{wyatt-dent-2002}.
For the $\sim 23\Myr$-old $\beta$~Pic disc, taking parameters from Table~\ref{tab:sample},
the same equations give $|\dot{M}| \approx 340 M_\oplus \Gyr^{-1} $ and
$M_0 \geq 31 M_\oplus$.

Alternatively, one can estimate the mass loss rate from the vertical thickness of the 
disc at (sub)mm wavelengths, which can nowadays be accurately measured for nearly 
edge-on discs with ALMA \citep{daley-et-al-2019,matra-et-al-2019b}. Indeed, ALMA 
observations show large, mm-sized dust, which is not succeptible to 
non-gravitational forces and thus should be co-located with the dust-producing 
planetesimals. Thus the vertical thickness of the disc directly measures orbital 
inclinations of planetesimals and, assuming energy equipartiton to relate inclinations to
eccentricities, also the relative velocities of colliding planetesimals.
Using standard formulae for collisional rates, 
we can then estimate the mass loss rate from the disc.
For the $\beta$~Pic disc,
\citet{matra-et-al-2019b} inferred mean planetesimal inclinations of $8.9^\circ$
for the dynamically hot population.
From this, we estimate the current mass loss rate of $\approx 170 M_\oplus \Gyr^{-1}$ 
and, from Eq.~(\ref{eq:M-Mdot simple}), an initial mass of $\geq 15 M_\oplus$, which is within
a factor of two consistent with the result obtained above with a different method.

A more detailed calculation of the minimum mass for a larger sample of discs is done later in the 
paper, in Sect.~\ref{ss:born small}.
However, these estimates already suggest that,
for bright debris discs,
a total disc mass of $\ga 10 M_\oplus$ is sufficient to collisionally sustain
the debris dust at the observed level even in Gyr-old systems.
We stress again that this is the minimum mass.
Since the disc likely also contains planetesimals that are too big to get collisionally disrupted
during its age, the actual disc mass will be larger. 

\subsection{The maximum mass of a debris disc}
\label{ss:max mass}

The maximum possible mass of a debris disc is determined by the total mass of condensible
compounds~-- refractories and volatiles~-- that were available in the
protoplanetary disc (PPD), out of which the debris disc and possibly planets formed.
We start with a simple estimate.
Assuming the PPD mass to be $0.1 M_\star$ \citep[e.g., since more massive discs would be gravitationally unstable,][]{haworth-et-al-2020} and using
a standard dust-to-gas ratio of $0.01$,
the maximum total mass in solids in the entire disc of a $1M_\odot$ star is
$0.001 M_\odot$, or $1 M_\mathrm{jup}$, or $300 M_\oplus$. 
Since planetesimals are made of this material, this is also the maximum total mass
that the debris disc emerging in the system after the gas dispersal can have.
The actual debris disc mass is expected to be much lower, because
a debris disc occupies a more or less narrow radial zone around the star.
While the outer edge of the disc probably marks the maximum distance at which 
planetesimals are able to form, the inner edge could be set by planets that
form in the debris disc cavity. Solids in the inner region will be
consumed to build rocky planets and the cores of giant planets, and
a fraction of solids may get ejected to interstellar space or fall down
to the star as a result of gravitational clearance by nascent planets or just gas drag.
This suggests that a more realistic estimate of the maximum debris disc mass could be
$\la 100 M_\oplus$.

However, this estimate is subject to additional uncertainties.
First of all, the PPD mass, which we assumed to be $0.1 M_\star$,
can in fact be a factor of several higher or lower.
Also, the masses of the observed protoplanetary discs have been found 
from (sub)mm observations to scale with
the stellar mass as $\propto M_\star^\alpha$, with $\alpha$ ranging from
$1.0$ \citep{williams-cieza-2011} to $2.7$ \citep{pascucci-et-al-2016}.
This means that the solid mass in the PPDs of more massive stars
(e.g. A-stars with 2 solar masses) can be higher by a factor $2$...$7$.
For instance, Fig. 5 in \citet{williams-cieza-2011} suggests the total
masses of PPDs around A stars to be between $2$ and $200 M_\mathrm{jup}$,
implying $6$--$600 M_\oplus$ in solids. The caveat of this comparison is that some 
additional solid mass can be hidden in larger bodies, planetesimals or even planetary 
cores, if they form early in the PPDs.

One more uncertainty is related to the assumed dust-to-gas ratio.
A debris disc is essentially made of ``metals'', i.e., all elements that are heavier 
than hydrogen and helium.  Some of the volatile compounds (such as water ice) also contain
hydrogen, yet the hydrogen mass in debris disc material is negligible. At the cold 
temperatures typical of extrasolar Kuiper belts, nearly all metal-based compounds,
both refractories and volatiles, will be in the solid state.
Thus the product of the mass fraction in metals $Z$ and the PPD mass should be a good proxy 
for the maximum  possible mass of a debris disc eventually emerging in the systems 
\citep{greaves-et-al-2007}.
Assuming that elemental abundances were originally the same in the young star and its disc, 
and that the abundances of all metals relative to iron are close to solar ones
\citep[e.g.,][]{lodders-2003}, we can estimate $Z$ from the stellar metallicity [Fe/H]
by using conversions such as
\citep{bertelli-et-al-1994}
\be
  \log Z  = 0.977 \mathrm{[Fe/H]} - 1.699.
\label{eq:metallicity conversion}
\ee
Taking $-0.5 \la \mathrm{[Fe/H]} \la 0.5$ as a typical range for debris disc host stars, 
this would give $0.006 \la Z \la 0.06$.

To get a handle on the absolute upper limit of the debris disc mass we can expect,
we can select a massive ($2 M_\odot$) A-type star
of a high metallicity ($\mathrm{[Fe/H]}=0.5$) and assume that the original PPD was also 
quite massive ($0.2 M_\star$). This would yield the maximum mass in solids of $8200 
M_\oplus$ (of which, however, only a fraction would reside in the annulus 
occupied by the debris disc).

Taking into account all the uncertainties discussed above,
we conclude that the total mass of a debris disc should not exceed
$\sim 100...1000$ Earth masses.
A general caveat of these estimates is that a gravitationally unstable disc
could be continually replenished so that the total mass available is larger
than that instantaneously available \citep[see, e.g.,][and references therein]{manara-et-al-2018}.

\subsection{Disc mass estimates assuming ideal collisional cascade}
\label{ss:ideal estimates}

\begin{figure}
\includegraphics[width=1.03\columnwidth]{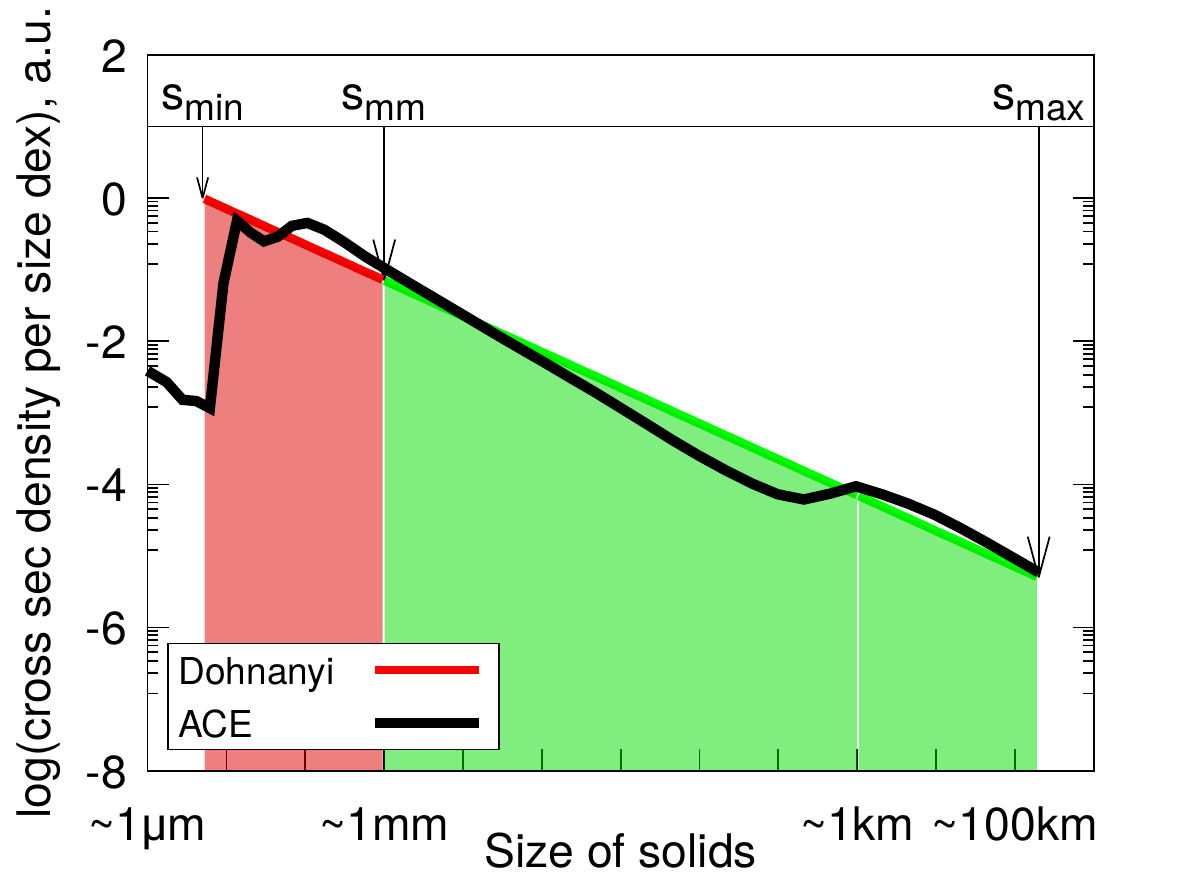}
    \caption{Schematic of an idealized size distribution.
     Plotted is the cross section per unit size decade as a function of object's radius.
     In these axes, a horizontal line corresponds to a power-law size distribution with $q=3$.
     Straight red-green line: the \citet{dohnanyi-1969} distribution with $q=3.5$.
     Black line: a typical outcome of a collisional simulation with the ACE code
     (run~A in Appendix~\ref{ss:ACE sims} after 100~Myr of evolution)
     that assumes
     a size-dependent critical fragmentation energy $Q_\mathrm{D}^*$ \citep[see Eq.~(1) in][]{krivov-et-al-2018}.
     Red-filled area shows what we refer to as ``dust sizes.''
     It corresponds approximately to the size range
     that is probed by debris disc observations from optical to (sub)mm wavelengths.}
    \label{fig:dohnanyi}
\end{figure}

To estimate the mass of a debris disc, we have to extrapolate the mass of the
observed dust to the mass of the unobservable dust-producing planetesimals.
This can be done with the help of debris disc models that tell us what the
size distribution of material from dust to the largest bodies may look like.
We first assume that the differential size distribution of material in the disc is
$n(s) \propto s^{-q}$ with $q=3.5$.
This is the slope expected in an infinite collisional cascade
with a constant strength of bodies
\citep{dohnanyi-1969}.

The Dohnanyi size distribution, on which we impose some lower and upper cut-offs discussed below,
is illustrated in Fig.~\ref{fig:dohnanyi}.
For comparison, we also show a typical size distribution from a collisional
simulation (run~A in Appendix~\ref{ss:ACE sims}),
in which a finite size interval was considered and 
the size dependence of the critical fragmentation energy was taken
into account.
These distributions are not very different.
The most tangible deviations are seen at dust sizes where radiation pressure
is important (red-filled area in Fig.~\ref{fig:dohnanyi}).
This causes a sharp cutoff below the blowout size \citep{burns-et-al-1979}
and some ripples above it 
\citep[e.g.,][]{campo-et-al-1994b,durda-dermott-1997,krivov-et-al-2006,thebault-augereau-2007,wyatt-et-al-2011}.
In some systems, dust grains can also be affected
by transport mechanisms (e.g., Poynting-Robertson and solar wind drag) which, in combination
with radiation pressure and collisions,
may cause further deviations of the slope from the Dohnanyi one
\citep[e.g.,][]{krivov-et-al-2006,thebault-wu-2008,krivov-2010,wyatt-et-al-2011,matthews-et-al-2013}.
At larger sizes, but below hundreds of metres, where the bodies are still kept together by molecular
forces rather than gravity (``strength regime''), the simulated size distribution
is somewhat steeper than the Dohnanyi $q=3.5$ \citep[e.g.,][]{durda-dermott-1997,o'brien-greenberg-2003}.
Still larger bodies whose strength is set by gravity (``gravity regime'')
exhibit a distribution that is flatter than the Dohnanyi one. 
Finally, the largest bodies, above about $1\km$ in radius, have collisional
lifetimes longer than the simulation time. As a result,
they preserve their initial size distribution which, in the simulation shown, was
assumed to have a slope of $3.7$.
Despite all these deviations from a single $q=3.5$ slope, Fig.~\ref{fig:dohnanyi} shows that
overall the Dohnanyi size distribution is a good approximation to more detailed models.

\tabcolsep 3pt
\begin{table*}
        \centering
        \caption{Sample of debris discs and their dust and disc mass estimates.
                }
        \label{tab:sample}
        \begin{tabular}{rlrlrrrrrrcccclr}
                \hline
HD 	&	Name	&$d$	&SpT	&$L_*$	    &$M_*$	  &Age	&$R$	&$\Delta R$	&$R_\mathrm{BB}$&$f_\mathrm{d}$ &$M_\mathrm{d}$ &$M_\mathrm{disc}$ &$M_\mathrm{disc}^*$ &
$s_\mathrm{km}$	& $M_\mathrm{disc}^\mathrm{min}$\\
  	&		&(pc)	&	&($L_\odot$)&($M_\odot$)  &(Myr)	&(au)	&(au)	&(au)		&		&($M_\oplus$)	&($M_\oplus$)	   &($M_\oplus$)	&
(km)		& ($M_\oplus$)\\
                \hline
$\cdots$& Kuiper belt	&10.0	&G2V	&1.0		&1.0		&4567	&45.0	&10.0		&20.0	&1.0e-07	&    6.7e-07 &   1.0e-02 &   $\cdots$ 	&$\cdots$ & $\cdots$\\
9672	&	49~Cet	&59.4	&A1V	&15.8		&1.9		&40	&228.0	&310.0		&85.4	&7.2e-04	&    2.8e-01 &   4.2e+03 &    3.6e+04 & 0.1  & 14.3 \\
15115	&	$\cdots$&45.2	&F4IV	&3.6		&1.3		&23	&78.2	&69.6		&55.1	&4.6e-04	&    8.5e-02 &   1.3e+03 &    2.7e+03 & 0.5  &  7.0 \\
21997	&	$\cdots$&71.9	&A3IV/V	&9.9		&1.7		&30	&106.0	&88.0		&65.4	&5.6e-04	&    9.8e-02 &   1.5e+03 &    4.2e+03 & 0.3  &  7.2 \\
22049	& $\varepsilon$~Eri&3.2	&K2Vk:	&0.3		&0.8		&600	&69.4	&11.4		&19.5	&4.0e-05	&    2.4e-03 &   3.6e+01 &    9.7e+01 & 0.3  &  0.2 \\
39060	& $\beta$~Pic	&19.4	&A6V	&8.1		&1.6		&23	&100.0	&100.0		&24.3	&2.1e-03	&    7.9e-02 &   1.2e+03 &    4.1e+03 & 0.3  &  5.5 \\
61005	&	$\cdots$&35.3	&G8Vk	&0.7		&0.9		&40	&66.4	&23.6		&21.0	&2.3e-03	&    1.3e-01 &   2.0e+03 &    2.0e+03 & 1.0  & 14.2 \\
95086	&	$\cdots$&90.4	&A8III	&6.1		&1.7		&15	&204.0	&176.0		&46.5	&1.4e-03	&    3.7e-01 &   5.6e+03 &    5.9e+04 & 0.1  & 17.7 \\
109085	&	$\eta$~Crv&18.3	&F2V	&5.0		&1.4		&1400	&152.0	&46.0		&52.9	&2.9e-05	&    3.1e-02 &   4.7e+02 &    8.1e+02 & 0.6  &  2.7 \\
111520	&	$\cdots$&108.6	&F5/6V	&3.0		&1.3		&15	&96.0	&90.0		&58.5	&1.1e-03	&    2.9e-01 &   4.4e+03 &    9.6e+03 & 0.4  & 23.6 \\
121617	&	$\cdots$&128.2	&A1V	&17.3		&1.9		&16	&82.5	&54.8		&30.0	&4.9e-03	&    2.8e-01 &   4.3e+03 &    4.6e+03 & 0.9  & 29.3 \\
131488	&	$\cdots$&147.7	&A1V	&13.1		&1.8		&16	&84.0	&44.0		&35.6	&2.2e-03	&    7.2e-01 &   1.1e+04 &    7.3e+03 & 1.6  & 85.8 \\
131835	&	$\cdots$&122.7	&A2IV	&11.4		&2.0		&16	&91.0	&140.0		&57.0	&2.2e-03	&    3.8e-01 &   5.8e+03 &    6.4e+03 & 0.9  & 39.1 \\
138813	&	$\cdots$&150.8	&A0V	&16.7		&2.2		&10	&105.0	&70.0		&69.6	&6.0e-04	&    4.3e-01 &   6.6e+03 &    1.2e+04 & 0.5  & 37.2 \\
145560	&	$\cdots$&133.7	&F5V	&3.2		&1.4		&16	&88.0	&70.0		&22.0	&2.1e-03	&    3.3e-01 &   4.9e+03 &    7.3e+03 & 0.7  & 30.0 \\
146181	&	$\cdots$&146.2	&F6V	&2.6		&1.3		&16	&93.0	&50.0		&17.0	&2.2e-03	&    1.7e-01 &   2.6e+03 &    6.9e+03 & 0.3  & 13.3 \\
146897	&	$\cdots$&128.4	&F2/3V	&3.1		&1.5		&10	&81.0	&50.0		&15.6	&8.2e-03	&    2.0e-01 &   3.0e+03 &    6.0e+03 & 0.5  & 16.7 \\
181327	&	$\cdots$&51.8	&F6V	&2.9		&1.3		&23	&86.0	&23.2		&50.1	&2.1e-03	&    3.2e-01 &   4.8e+03 &    5.8e+03 & 0.8  & 31.3 \\
197481	&	AU~Mic	&9.9	&M1Ve	&0.1		&0.6		&23	&24.6	&31.6		&11.9	&3.3e-04	&    1.4e-02 &   2.2e+02 &    1.2e+02 & 1.9  &  1.8 \\
216956	&	Fomalhaut&7.7	&A4V	&16.1		&1.9		&440	&143.1	&13.6		&72.2	&7.5e-05	&    2.4e-02 &   3.6e+02 &    9.6e+02 & 0.3  &  1.8 \\
218396	&	HR~8799	&39.4	&F0V	&5.5		&1.5		&30	&287.0	&284.0		&123.6	&2.5e-04	&    1.7e-01 &   2.6e+03 &    7.5e+04 & 0.02 &  5.9 \\
                \hline
        \end{tabular}

{\em Notes:} See Table~1 in \citet{matra-et-al-2018} for references to the stellar and disc parameters.
$M_\mathrm{d}$ is the dust mass estimated from sub-mm fluxes (Eq.~\ref{eq:dust mass}).
$M_\mathrm{disc}$ is the total disc mass, obtained by extrapolation of the dust mass
assuming two different power laws
between $s_\mathrm{mm} = 1\mm$ and $s_\mathrm{km} = 1\km$
and between $s_\mathrm{km}=1\km$ and $s_\mathrm{max}=200\km$
(Eq.~\ref{eq:disc mass}).
$M_\mathrm{disc}^*$ is the total disc mass 
corrected for the time-dependent transition size $s_\mathrm{km}$ (Eq.~\ref{eq:corrected disc mass final}).
$M_\mathrm{disc}^\mathrm{min}$ is the total disc mass up to that size $s_\mathrm{km}$ (Eq.~\ref{eq:min disc mass}).
\end{table*}

The size distribution in a debris disc has several characteristic sizes.
The lower cutoff of the size distribution, $s_\mathrm{min}$, is close to the radiation pressure
blowout limit \citep[e.g.,][]{pawellek-krivov-2015}.
This is generally on the order of $1\mum$, with the exact value depending on
the luminosity and mass of the central star and the dust composition
\citep[e.g.,][]{kirchschlager-wolf-2013}.
Another characteristic size, $s_\mathrm{mm}$, is the maximum size of particles that we refer to as
``dust.''
A commonplace choice, which we also make here, is to set $s_\mathrm{mm} = 1\mm$,
so that it corresponds to the radius of the particles that are most efficiently probed
by (sub)mm observations of disc thermal emission.
Also, this is the size of the grains above which non-gravitational effects such a radiation pressure
can be assumed to be negligible.
As explained above, the details of the size distribution between $s_\mathrm{min}$
and $s_\mathrm{mm}$ are still not fully understood. Therefore, in this paper
we circumvent potential uncertainties by working with dust masses derived from (sub)mm thermal
emission fluxes (with a caveat that the size distribution at dust sizes indirectly affects conversion
of fluxes to dust masses, see Sect.~\ref{ss:opacity}).

Finally, there is an upper cutoff to the size distribution at $s_\mathrm{max}$.
In this work we consider a nominal value of $200\km$, but this is varied in later sections.
The justification for this choice is two-fold.
First, this value is close to the radius of the largest objects in the cold classical Kuiper belt,
although the size distribution in the Kuiper belt steepens to values $\gg 4$ already above
$\sim 50$--$100\km$ \citep[e.g.,][]{bernstein-et-al-2004}.
Second, this is about the size of the largest planetesimals that are built in
state-of-the-art planetesimal formation models prior to the onset of runaway growth or pebble accretion.
This equally applies to models based on incremental collisional growth 
\citep[$50$--$100\km$; see, e.g.,][]{kobayashi-et-al-2016} and those invoking pebble concentration
\citep[$80$--$600\km$; see, e.g.,][]{schaefer-et-al-2017}.

As the size distribution slope is less than $4$,
the mass of the disc $M_\mathrm{disc}$ is dominated by
the largest bodies of size $s_\mathrm{max}$:
$M_\mathrm{disc} \propto s_\mathrm{max}^{4-q}$.
This means that the total disc mass in bodies, say, up to $200\km$ is radius, must
be higher than the dust mass in mm-sized particles, $M_\mathrm{d}$
(which are probed by sub-mm observations), by 4 orders of magnitude
(for $q=3.5$).
Taking $M_\mathrm{d} \sim 10^{-1} M_\oplus$, as appropriate for bright debris discs
\citep[see, e.g., Fig.~34 in][]{holland-et-al-2017},
gives $M_\mathrm{disc} \sim 1000 M_\oplus$, which is close to the
the maximum possible mass of a debris disc, as discussed above. 

\begin{figure}
\includegraphics[width=\columnwidth]{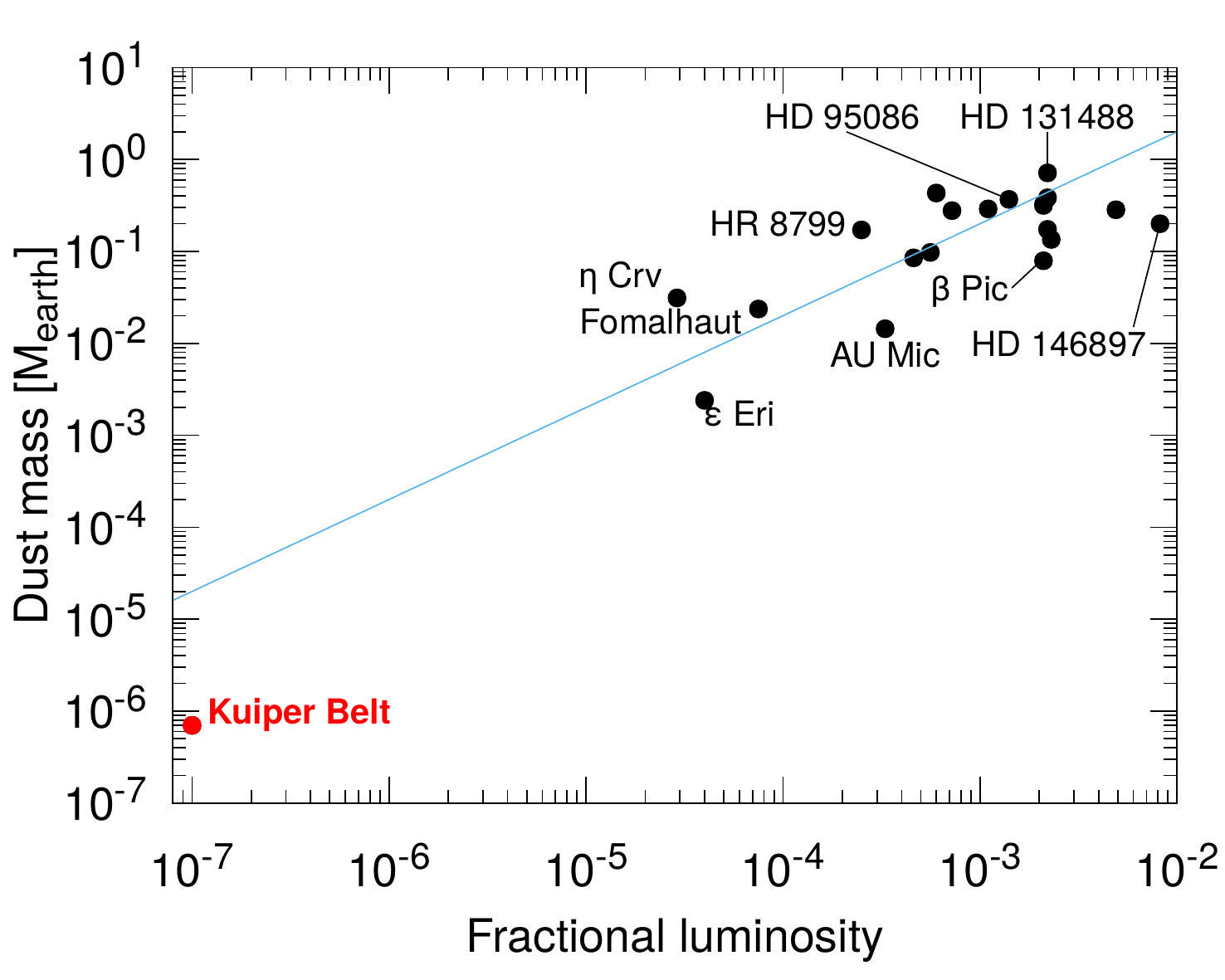}
    \caption{Dust masses of debris discs in our sample.
       Thin line is the expected $M_\mathrm{d} \propto f_\mathrm{d}$ dependence.
       Some of the prominent discs are labeled.
       Red circle shows the Solar system's Kuiper belt for comparison.
       }
    \label{fig:dustmass-fd_slide}
\end{figure}

To get more accurate estimates of the debris disc mass,
we now select a set of well-known debris discs. To this end, we take a sample
of discs for which resolved images have been taken with ALMA or SMA from
\citet{matra-et-al-2018}.
Their Table~1 lists the luminosity $L_\star$, mass $M_\star$, and an age estimate
of the central star, as well as the characteristic radius $R$ and radial extent $\Delta R$
of each disc.
A number of other quantities useful for the analysis, such as
the fractional luminosity $f_\mathrm{d}$ of each disc,
are also given.
Finally, total (sub)mm fluxes for each of the discs, $F_\nu$, are available.
From these, it is easy to calculate the dust mass (i.e., the mass of grains with
radii from $s_\mathrm{min}$ to $s_\mathrm{mm} = 1\mm$):
\be
  M_\mathrm{d}
  \approx
  {
    F_\nu d^2
    \over
    \kappa B_\nu(T_\mathrm{d})
  } ,
\label{eq:dust mass}
\ee
where $d$ is the distance to the disc host star, 
$\kappa$ is the dust opacity (which we assume to be $1.7\cm^{2}\g^{-1} (850\mum/\lambda)$
for consistency with previous studies, e.g.,
\citeauthor{zuckerman-becklin-1993}
\citeyear{zuckerman-becklin-1993};
\citeauthor{holland-et-al-1998}
\citeyear{holland-et-al-1998};
\citeauthor{holland-et-al-2017}
\citeyear{holland-et-al-2017}),
and $B_\nu$ is the black body emission intensity for the dust temperature $T_\mathrm{d}$.
The latter can be inferred from
a disc's ``black body'' radius $R_\mathrm{BB}$,
i.e., the radius that a disc of a given temperature (consistent with the observed spectral
energy distribution, SED) would have if its grains were absorbing and emitting as black bodies:
$T_\mathrm{d} = 278\K (L_*/L_\odot)^{1/4} (R_\mathrm{BB}/\AU)^{-1/2}$.
The total disc mass in bodies up to the radius $s_\mathrm{max}$
is calculated as
\be
 M_\mathrm{disc}
 = 
 M_\mathrm{d}
 \left(s_\mathrm{max} \over s_\mathrm{mm} \right)^{4-q} .
\label{eq:disc mass dohnanyi}
\ee

\begin{figure}
\includegraphics[width=\columnwidth]{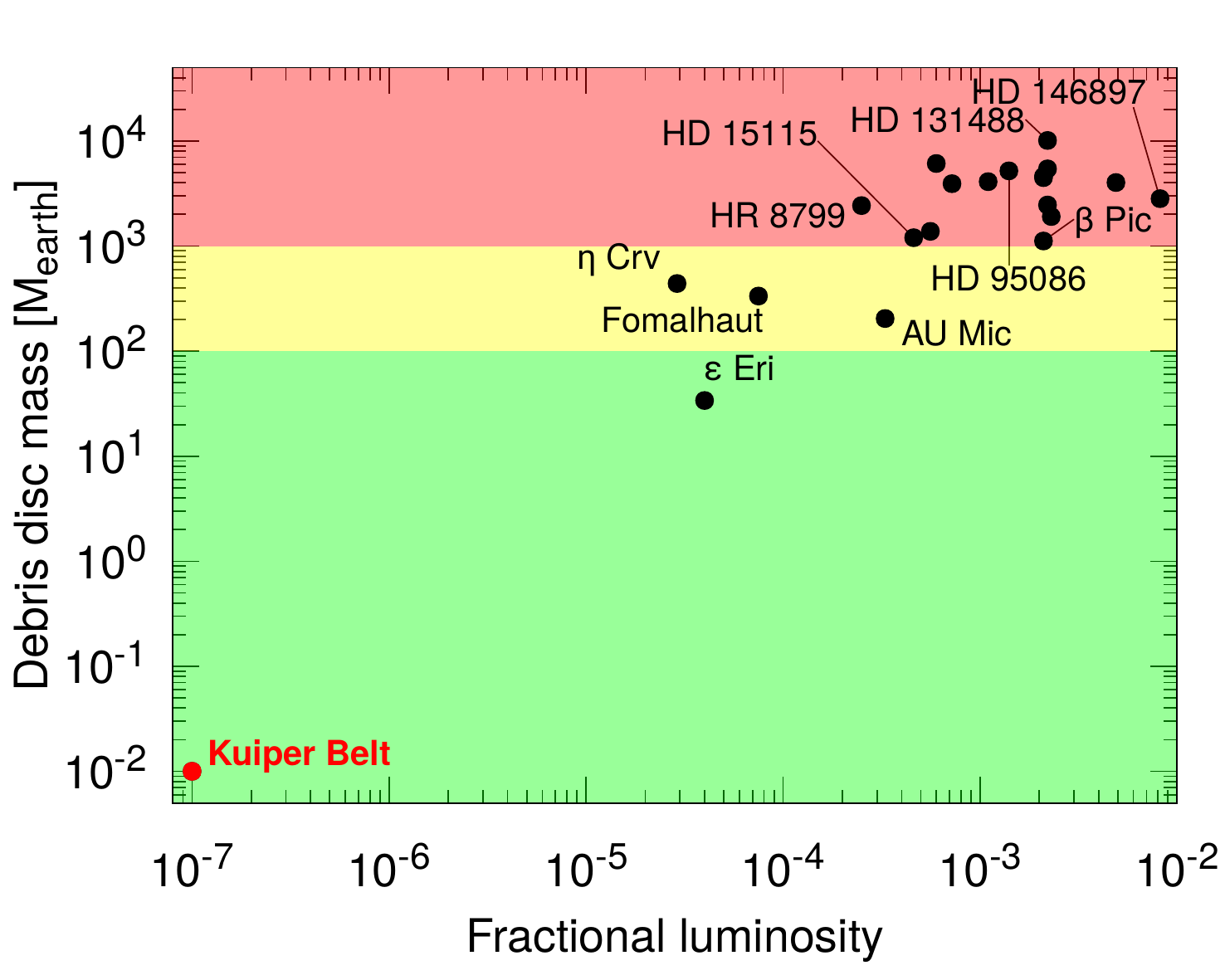}\\
\includegraphics[width=\columnwidth]{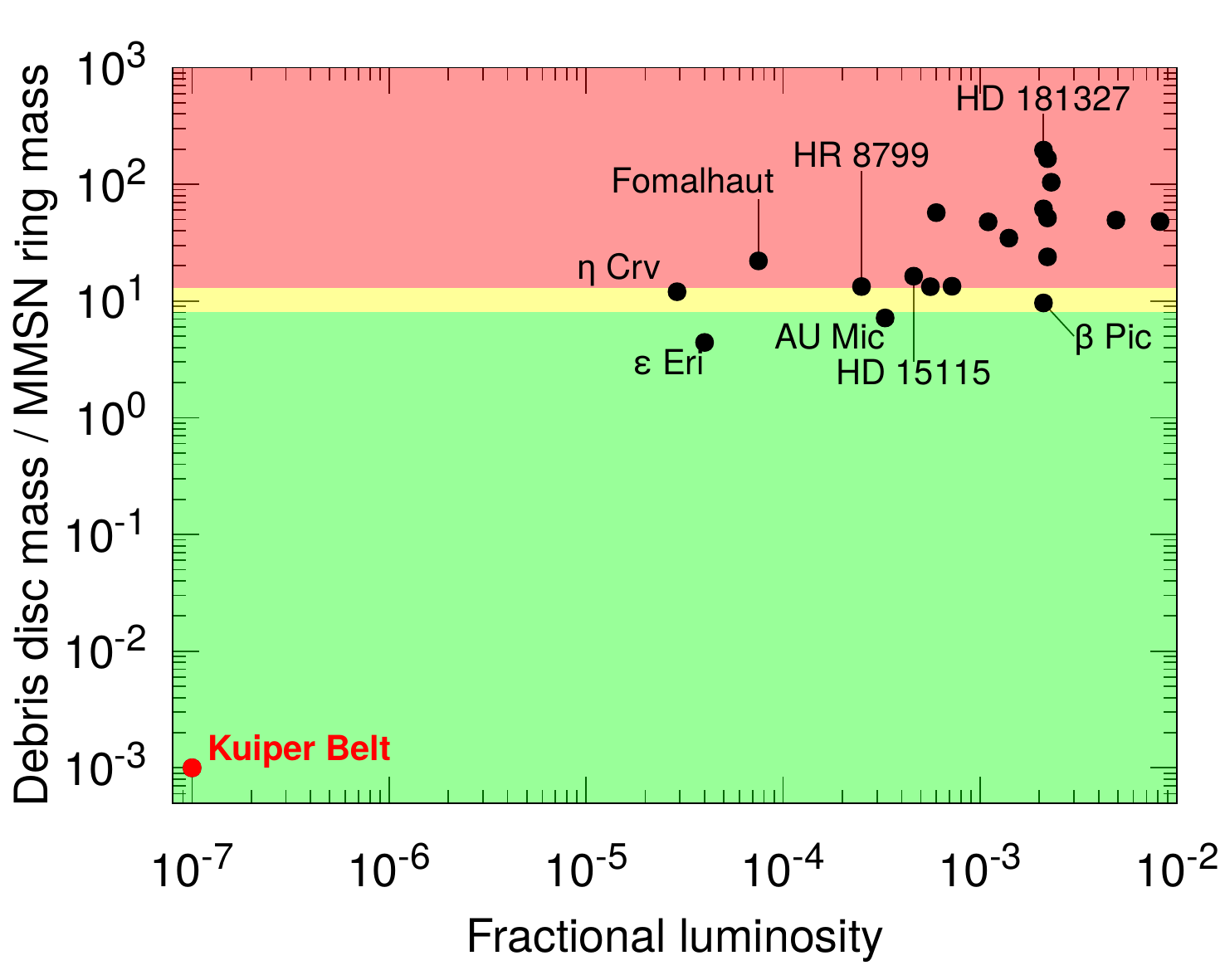}
    \caption{Total mass for the same selection of debris discs, estimated
       under the assumption of the Dohnanyi collisional cascade
       ($q=3.5$) in the size range from $1\mm$ to $200\km$.
       Top: disc masses in the units of Earth masses.
       Bottom: disc masses in the ``MMSN units.''
       Traffic light-style background colors mark the allowed (green) and forbidden (red) areas,
       and a boundary between them, corresponding the the ``maximum debris disc mass'' (yellow).
       }
    \label{fig:mass&xm-fd_3.5_1.0mm_3.5_1.0km_3.5_200km_slide}
\end{figure}

Of the 26 sources in \citet{matra-et-al-2018},
we selected 20 with best-quality data (L. Matr{\`a}, pers. comm.).
All of these were observed with ALMA in bands~6 ($\lambda \approx 1.3\mm$)
or~7 ($\lambda \approx 850\mum$).
As a reference, to this sample we added the 
Kuiper belt of our Solar system with
$R = 45\AU$,
$\Delta R = 10\AU$,
$f_\mathrm{d} = 10^{-7}$ \citep{vitense-et-al-2012}.
To obtain an estimate of the dust mass in the Kuiper belt in the same way as for extrasolar discs,
we need a dust flux at some sub-mm wavelength that would be detected from the Kuiper belt
if it were observed from afar.
That wavelength and the distance at which the Kuiper belt is placed can be taken arbitrarily, and
we use $F_\nu = 1.48\mu\mathrm{Jy}$ at $850\mum$,
as predicted for the Kuiper belt if seen from a 10~pc distance 
\citep[see Fig.~8 in][]{vitense-et-al-2012}.
Note that, since the thermal emission of the Kuiper belt has not been detected,  the fractional luminosity
and the emission flux are model-dependent and are uncertain
by a factor of a few \citep[see, e.g.,][and references therein]{poppe-et-al-2019}.
However, this uncertainty is unimportant for our purposes, as we only wish to demonstrate that
the Kuiper belt is not affected by the mass problem.

The discs of the resulting sample and their essential parameters are listed in 
Table~\ref{tab:sample}.
The resulting dust masses from Eq.~(\ref{eq:dust mass}) are also given in
Table~\ref{tab:sample} and shown
in Fig.~\ref{fig:dustmass-fd_slide} against dust fractional luminosity.
As expected there is an approximately linear correlation between these two quantitites.

Applying now Eq.~(\ref{eq:disc mass dohnanyi}) with $q=3.5$
yields the top panel in Fig.~\ref{fig:mass&xm-fd_3.5_1.0mm_3.5_1.0km_3.5_200km_slide}.
The mass predicted for the Kuiper belt from its dust mass is low ($\sim 0.01 M_\oplus$), but pretty much consistent
with the values inferred in detailed analyses of TNO surveys
\citep[e.g.,][]{fraser-et-al-2014}.
The masses of all extrasolar discs in the sample are
by 3--6 orders of magnitude higher and increase with the fractional luminosity
as they should from Eq.~(\ref{eq:dust mass}).
While for some of the discs
($\varepsilon$~Eri, AU~Mic, Fomalhaut, $\eta$~Crv) the total mass lies between
$50$--$1000 M_\oplus$, the other discs appear to be unfeasibly massive.
The largest values we obtain are as large as
$M_\mathrm{disc} \sim 10^4 M_\oplus$!

We can also view these results in a different way, measuring the disc masses
in the units of the Minimum Mass Solar Nebula
\citep[MMSN;][]{weidenschilling-1977b,hayashi-1981}.
Following \citet{kenyon-bromley-2008}, we adopt the surface density of solids
in the MMSN in the form
\be
 \Sigma(r)
 =
 \Sigma_0 \; r^{-3/2} ,
\label{eq:Sigma_mmsn}
\ee
where $\Sigma_0 = 29.6 \g\cm^{-2}(M_\star/M_\odot)$
and the distance from the star $r$ is measured in $\AU$.
Note that this MMSN definition includes $M_\star$ to account for
the fact that the protoplanetary disc masses are
roughly proportional to the stellar masses
\citep[e.g.,][]{williams-cieza-2011}.
We then compute the mass that a debris disc with a radius $R$ and width $\Delta R$
(measured in $\AU$) would have if it had the MMSN surface density profile,
\be
 M_\mathrm{MMSN}
 =
 4\pi \Sigma_0
 \left[ 
  \left( R + \Delta R /2  \right)^{1/2}
  -
  \left( R - \Delta R /2  \right)^{1/2}
 \right] ,
\label{eq:M_mmsn}
\ee
and define $x_\mathrm{m} \equiv M_\mathrm{disc} / M_\mathrm{MMSN}$
to be the disc mass ``in MMSN units.''

The result is shown in the bottom panel
in Fig.~\ref{fig:mass&xm-fd_3.5_1.0mm_3.5_1.0km_3.5_200km_slide}.
Most of the discs in our sample would have $x_\mathrm{m}$ greater than 10
(up to $\sim 200$).
However, only surface densities of PPDs below $\sim 10$ times MMSN are commonly considered
plausible.
One reason for this is that higher densities would violate
the \citet{toomre-1964} gravitational stability criterion \citep[e.g.,][]{kuchner-2004,rafikov-2005},
while the vast majority of PPDs observed in young stellar associations
were found to be gravitationally stable at all radii
\citep[e.g.,][]{isella-et-al-2009}.

One caveat of measuring the disc masses in the MMSN units
is that the $-3/2$ radial slope in the MMSN definition of equation
(\ref{eq:Sigma_mmsn}) does not necessarily represent actual conditions in extrasolar systems.
Indeed, slopes of PPDs inferred from (sub)mm obsrvations are generally flatter
\citep[e.g.,][]{andrews-et-al-2009,andrews-et-al-2010}, whereas ``Minimum-Mass Extrasolar Nebulae''
constructed from known extrasolar planet populations have steeper surface densities
\citep{kuchner-2004,chiang-laughlin-2013}.

\subsection{Disc mass estimates including planetesimal formation model predictions}
\label{ss:formation}

The idealized models described above just consider a size distribution that is set by
a collisional cascade.
To arrive at a more realistic model, we must take into account that the largest planetesimals,
especially in younger systems, are not part of the cascade, because their collisional
lifetime may be longer than the system's age. Such planetesimals preserve their
original size distribution they acquired at the formation stage.
Although contributing only little to the production of the observed dust, these
are nevertheless important for understanding the mass budget, since they
contain most of the debris disc mass.

\begin{figure}
\includegraphics[width=1.03\columnwidth]{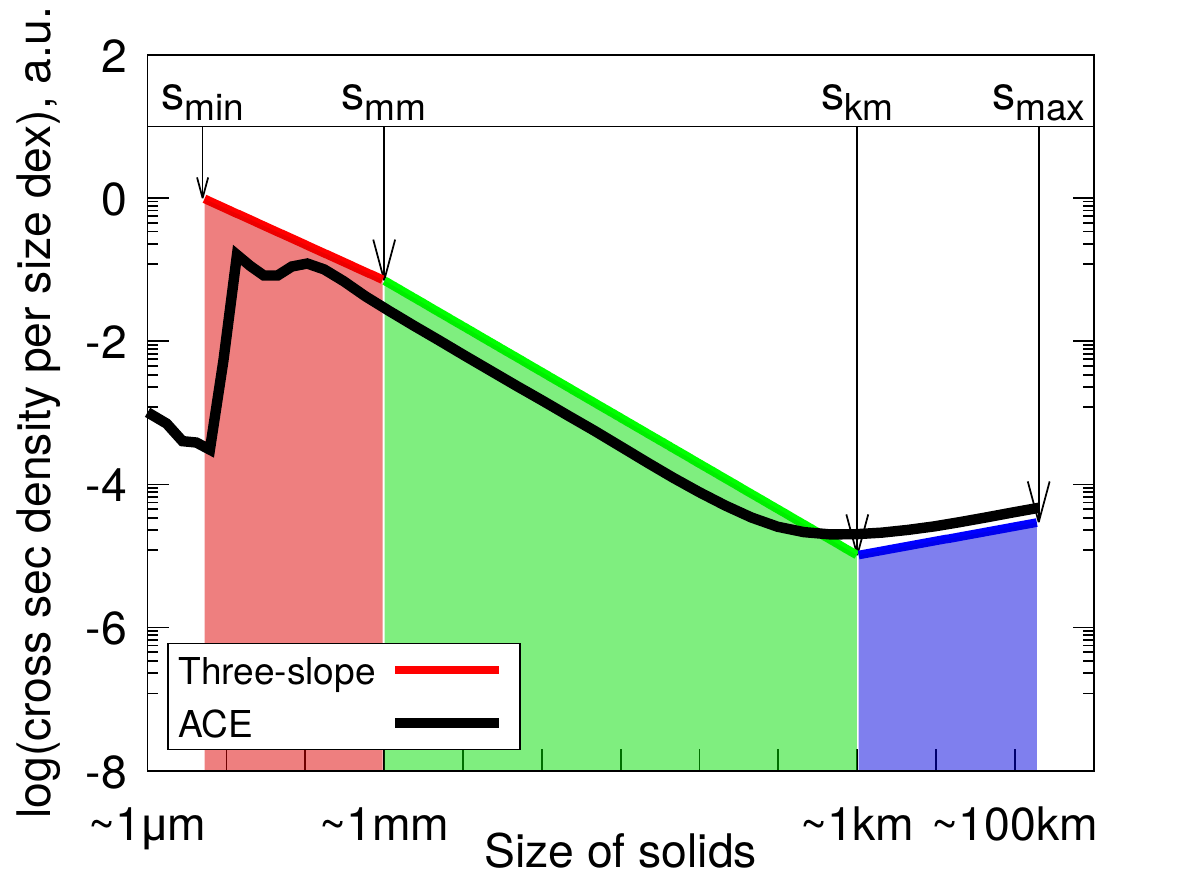}
    \caption{A more realistic size distribution.
     Three-segment (red-green-blue) line:
     a combination of three power laws with different slopes
     that are expected at the dust sizes (red-filled area), sizes of planetesimals that are collisionally
     ground by the cascade (green), at sizes of largest planetesimals that retain their primordial
     size distribution (blue).
     Black line: a typical outcome of a collisional simulation with the ACE code
     (run~B in Appendix~\ref{ss:ACE sims} after 100~Myr of evolution)
     that takes a primordial size distribution
     of planetesimals predicted by pebble concentration models of their formation.
     }
    \label{fig:pebble-piles}
\end{figure}

To understand what the primordial size distribution of large planetesimals might look like,
we have to take into account the planetesimal formation history.
Although several competing models exist, in this paper we consider the possibility that
planetesimals form
by the ``particle concentration'', or ``pebble concentration'' mechanism
\citep[e.g.,][]{haghighipour-boss-2003,johansen-et-al-2007,cuzzi-et-al-2008,cuzzi-et-al-2010,%
chambers-2010,johansen-et-al-2015,simon-et-al-2016,simon-et-al-2017}.
In this mechanism, pebble-sized (typically mm--cm sized) particles in a PPD concentrate
in certain zones of the disc, for instance by streaming instability (SI), followed by
a local gravitational collapse to form planetesimals in just 3--5~Myr
\citep[e.g.,][]{johansen-et-al-2015,carrera-et-al-2017}.

These models make predictions for the sizes of newly formed planetesimals.
The size distribution above $s \sim 1\km$ should be much flatter than the Dohnanyi one,
$q_\mathrm{big}=2.8 \pm 0.1$ for a broad range of possible model parameters
\citep{johansen-et-al-2015,simon-et-al-2016,simon-et-al-2017,abod-et-al-2018}.
This slope should extend to radii of a few hundred kilometres
(with $s_\mathrm{max} = 200\km$ being a typical value) before it steepens towards
Pluto-sized dwarf planets \citep{schaefer-et-al-2017}.

\begin{figure}
\includegraphics[width=\columnwidth]{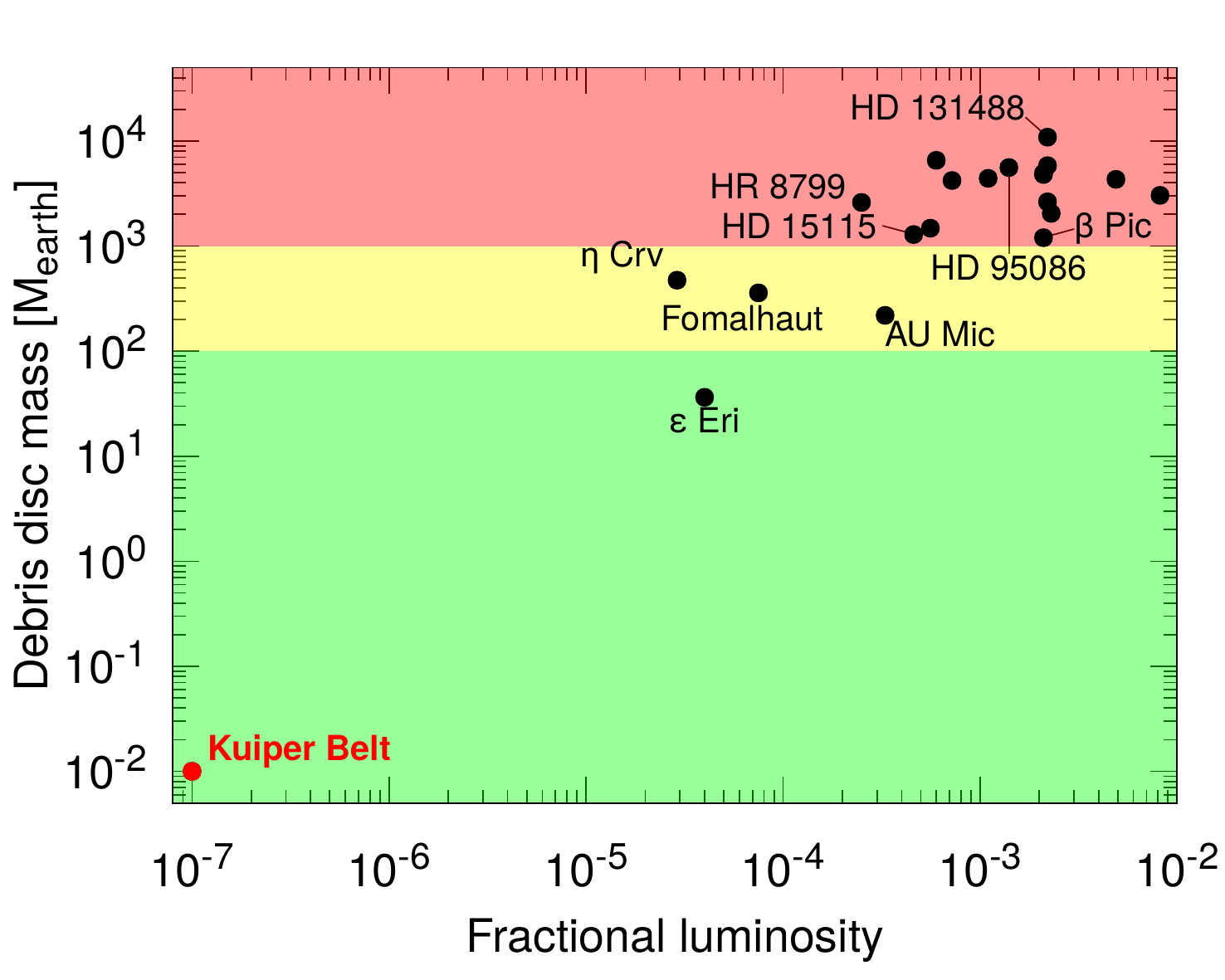}
    \caption{Same as Fig.~\ref{fig:mass&xm-fd_3.5_1.0mm_3.5_1.0km_3.5_200km_slide} top,
       but assuming that planetesimals form by the streaming instability (SI). 
       This incorporates the ``initial mass function'' of bodies predicted by the SI-based models
       of their formation with $q_\mathrm{big}=2.8$ up to a size of $200\km$.
       }
    \label{fig:mass&xm-fd_3.5_1.0mm_3.7_1.0km_2.8_200km_slide}
\end{figure}

The size distribution expected from the collisional disruption of planetesimals with such an initial size distribution
can be approximated by a combination of three power laws
(Fig.~\ref{fig:pebble-piles}).
One would be valid at dust sizes, from the blowout limit to some size $s_\mathrm{mm} \sim 1\mm$,
where radiation pressure is important.
As a nominal value, we take $q = 3.5$ (see a discussion in Sect.~\ref{ss:damping at mm}).
This slope $q$ can further be altered by transport mechanisms, such as Poynting-Robertson
and stellar wind drag.
Although these are usually of importance for low-mass discs that are not considered here,
they may also affect discs of late-type stars with strong winds.
Two discs in our sample, $\varepsilon$~Eri \citep{reidemeister-et-al-2011} 
and AU~Mic \citep{schueppler-et-al-2015}, belong to this category.
As explained in Sect.~\ref{ss:ideal estimates},
this slope does not affect the results of our analysis considerably (except that it contributes
to opacity, see Sect.~\ref{ss:opacity},
and slightly affects the extrapolation of the dust mass to the disc mass,
entering Eq.~\ref{eq:disc mass} below).
However, it is included 
to acknowledge the uncertainty that is inevitable when considering the quantity
of small dust that may be present, e.g., as probed in infrared and scattered light observations.

The second slope $q_\mathrm{med}$, extending
from $s_\mathrm{mm} \sim 1\mm$
to $s_\mathrm{km} \sim 1\km$,
is set by the collisional cascade and is somewhat steeper than the Dohnanyi one.
We fix it to $q_\mathrm{med} = 3.7$.
Finally, the third slope $q_\mathrm{big}$ is expected for the bodies larger than 
$s_\mathrm{km}$, which are not part of the 
cascade.
As discussed above, it should be the primordial one predicted by planetesimal formation models, 
i.e., $q_\mathrm{big} \approx 2.8$.
Figure~\ref{fig:pebble-piles} shows that the analytical three-phase size distribution with slopes
$q$, $q_\mathrm{med}$ and $q_\mathrm{big}$
agrees well with the collisional simulation.

We now apply this three-slope size distribution to the same sample of debris discs.
Equation~(\ref{eq:disc mass dohnanyi})
is replaced by
\be
 M_\mathrm{disc}
 = 
 M_\mathrm{d}
 \;
 {4 - q\phantom{_big} \over 4 - q_\mathrm{big}}
 \left(s_\mathrm{km} \over s_\mathrm{mm} \right)^{4-q_\mathrm{med}}
 \left(s_\mathrm{max} \over s_\mathrm{km} \right)^{4-q_\mathrm{big}} .
\label{eq:disc mass}
\ee

Disc masses obtained with Eqs.~(\ref{eq:dust mass}) and (\ref{eq:disc mass})
are given in the $M_\mathrm{disc}$ column of Table~\ref{tab:sample} and
plotted in Fig.~\ref{fig:mass&xm-fd_3.5_1.0mm_3.7_1.0km_2.8_200km_slide}.
The problem gets even worse, because there is now a lot of ``dead'' mass in bodies larger than
$\sim 1\km$. These big objects are not needed to reproduce the debris disc data with 
collisional models, although they are expected to be present 
from the planetesimal formation simulations.
However, a steeper slope $q_\mathrm{med}$ in Eq.~(\ref{eq:dust mass})
compared to $q$ in Eq.~(\ref{eq:disc mass dohnanyi}) reduces the total mass.
As a result, the disc masses shown in 
Fig.~\ref{fig:mass&xm-fd_3.5_1.0mm_3.7_1.0km_2.8_200km_slide}
are only 7\% larger than those depicted in
Fig.~\ref{fig:mass&xm-fd_3.5_1.0mm_3.5_1.0km_3.5_200km_slide}.

Equation~(\ref{eq:disc mass}) shows that the disc mass estimate
$M_\mathrm{disc}$ is proportional to 
$s_\mathrm{km}^{q_\mathrm{big} - q_\mathrm{med}}
\propto
s_\mathrm{km}^{-0.9}$.
Thus any uncertainty in $s_\mathrm{km}$ would translate to a similar uncertainty in
$M_\mathrm{disc}$.
To create Fig.~\ref{fig:mass&xm-fd_3.5_1.0mm_3.7_1.0km_2.8_200km_slide}, we simply
fixed $s_\mathrm{km}$ to $1\km$.
However, this size that separates the primodial
planetesimal population from the collisionally processed ones increases with time,
as larger and larger bodies get involved in the cascade.
The size dependence of $s_\mathrm{km}$ on age is discussed in more detail in
Sect.~\ref{ss:recent ignition}, where also quantitative estimates are given.
We will particularly show that using a time-variable $s_\mathrm{km}$ in
Eq.~(\ref{eq:disc mass}) would not significantly affect the mass estimate for any discs,
regardless of their age.

\begin{figure*}
\includegraphics[width=\columnwidth]{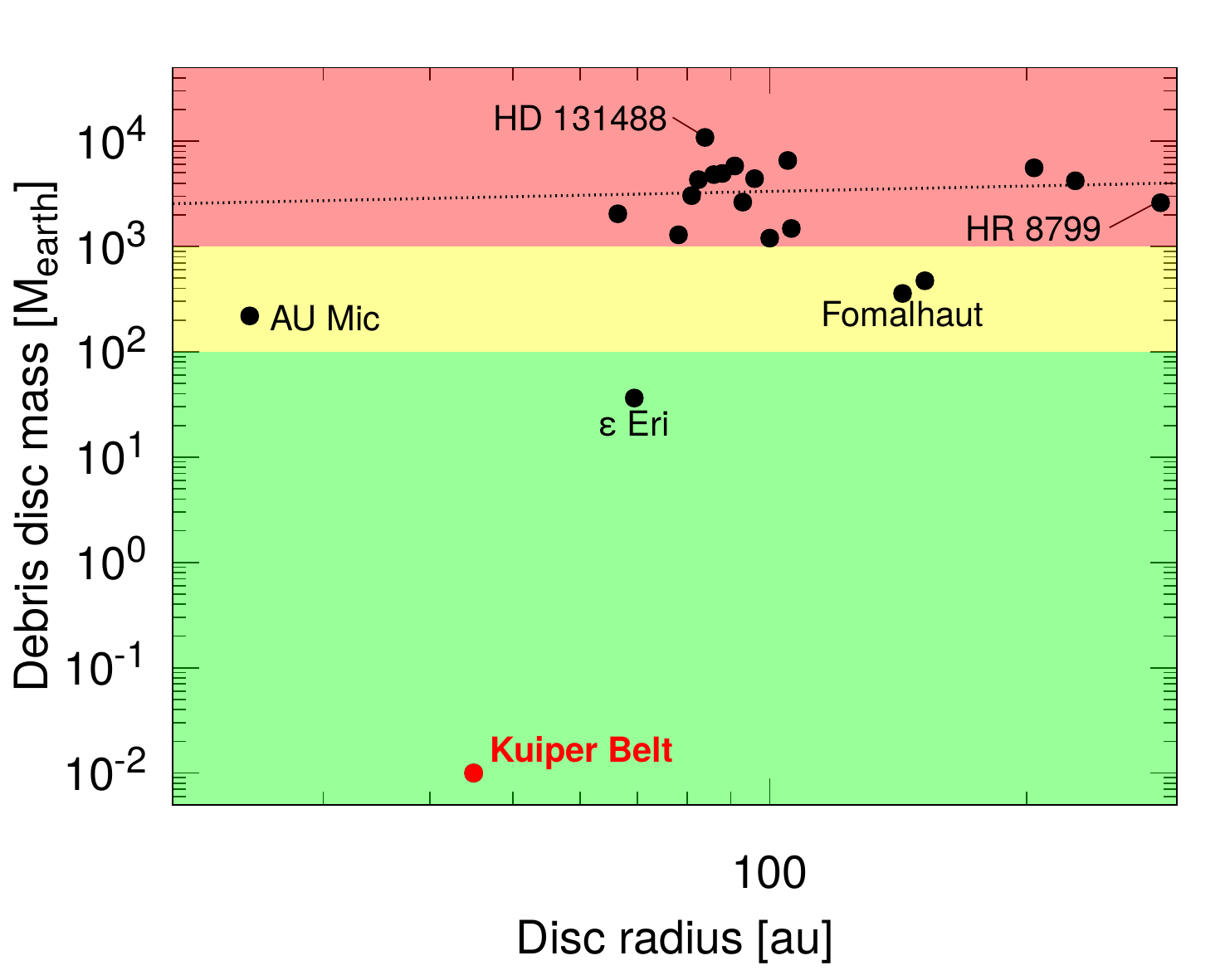}
\includegraphics[width=\columnwidth]{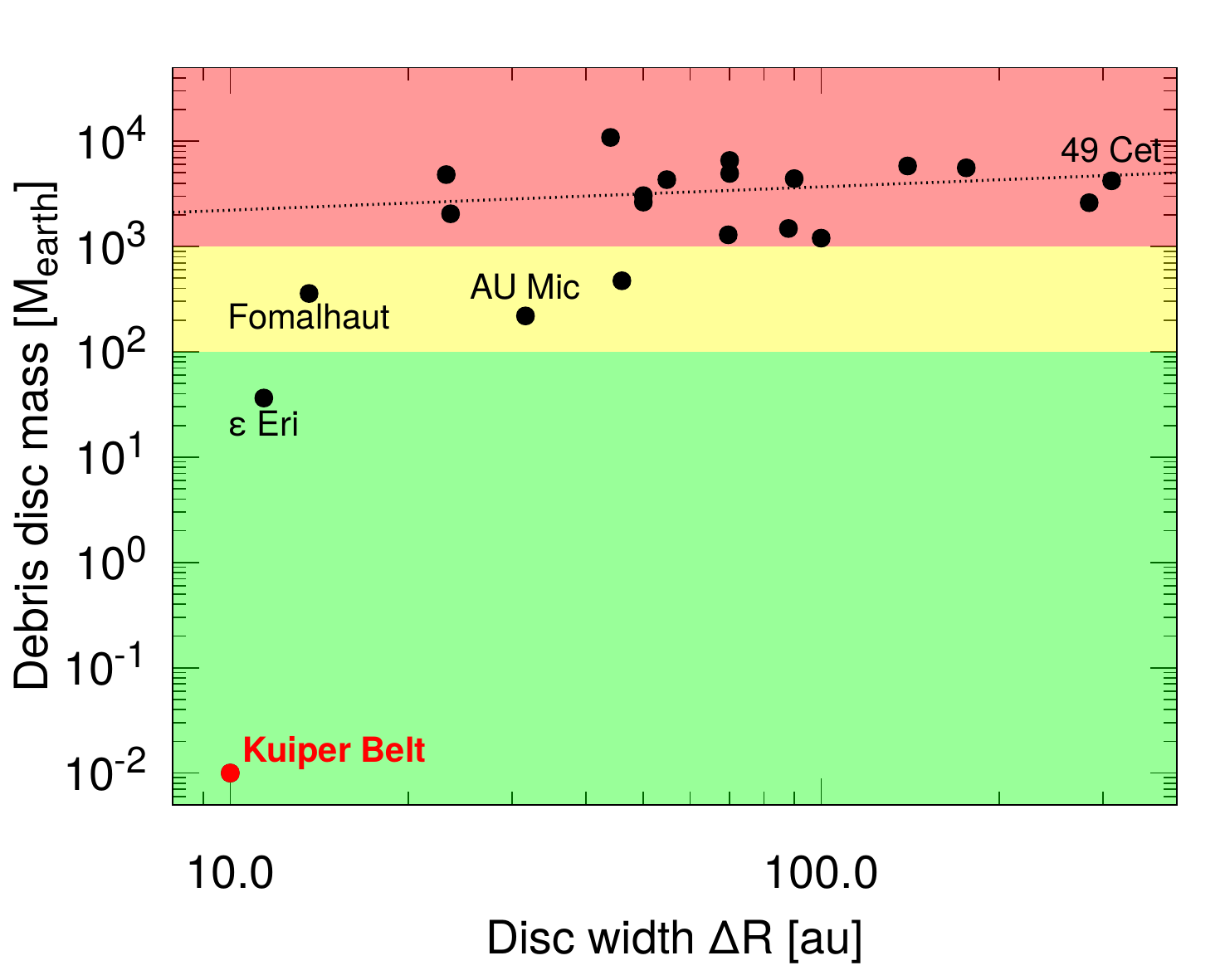}\\
\includegraphics[width=\columnwidth]{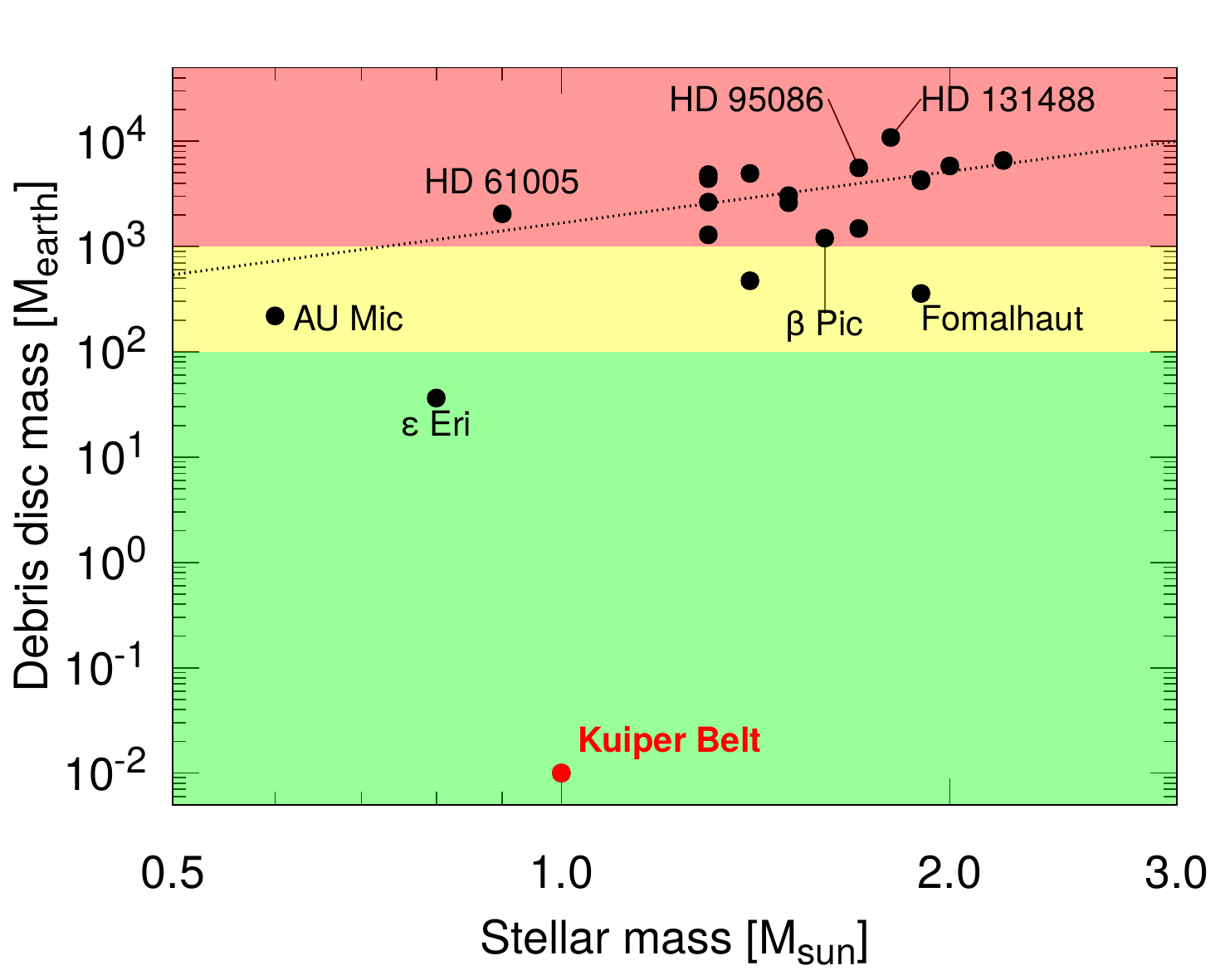}
\includegraphics[width=\columnwidth]{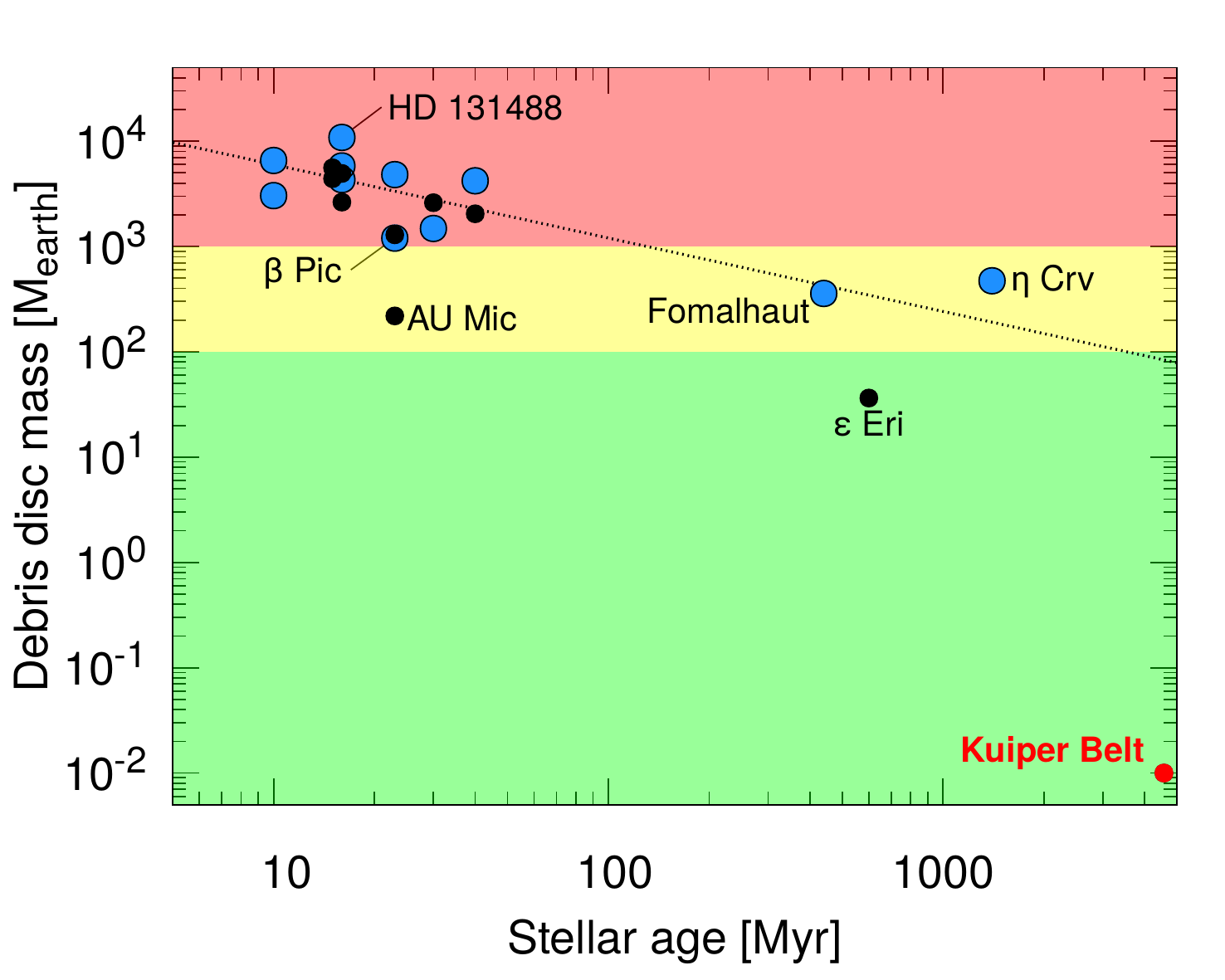}
    \caption{Dependence of the inferred total disc mass on
     (a)~the disc radius,
     (b)~the disc width,
     (c)~the stellar mass, and
     (d)~the stellar age.
     Dashed lines in each plot are best power-law fits through the data (without the Kuiper belt).   
     In panel~(d), larger blue symbols mark discs in which gas detections have been reported, whereas
     black symbols stand for discs without such detections.
     Otherwise, symbols, labels, and filled areas are as in
     Figs.~\ref{fig:mass&xm-fd_3.5_1.0mm_3.5_1.0km_3.5_200km_slide}
     and~\ref{fig:mass&xm-fd_3.5_1.0mm_3.7_1.0km_2.8_200km_slide}.
    }
    \label{fig:mass vs various}
\end{figure*}

In principle, pebble concentration models of planetesimal formation
have more implications than
just introducing a primordial population of planetesimals with a flat size distribution.
The same models predict the planetesimals to be loosely bound, fragile ``pebble piles''.
This prediction is backed up by a bulk of evidence that comets
in the Solar system are porous ``pebble piles'' \citep[e.g.,][]{blum-et-al-2017}, which probably
traces back to their formation mechanisms.
Similarly, (non-differentiated) asteroids \citep[e.g.,][]{britt-et-al-2002}
are known to have a significant macro-porosity and are thought
to have been turned into rubble piles by repeated
shattering collisions with subsequent gravitational reaccumulation.
As a result,
small bodies in the Solar system have relatively low densities.
For instance, a density of 
$0.533\g\cm^{-3}$ was measured for the comet 67/P \citep{paetzold-et-al-2016}.
Similar values were determined for Arrokoth \citep{stern-et-al-2019}, and densities in the
$0.5 \ldots 1\g\cm^{-3}$ range were inferred for a suite of Kuiper-belt objects up to hundreds of km in size
\citep[e.g.,][and references therein]{brown-2012}.

By analogy with the Solar system, we may expect large planetesimals in extrasolar
systems to have  low densities as well. To account for this, we can allow solids
in the three size ranges of the broken power-law
size distribution to have different bulk densities ($\rho$, $\rho_\mathrm{med}$, and $\rho_\mathrm{big}$).
In this case, Eq.~(\ref{eq:disc mass}) needs to be generalized.
This generalisation depends on how exactly the three-slope distribution is defined.
If we require that the differential {\em size} distribution is continuous at transition points
$s_\mathrm{mm}$ and $s_\mathrm{km}$ where the density changes, then
the right-hand side of Eq.~(\ref{eq:disc mass}) will acquire an additional factor
$\rho_\mathrm{big} / \rho$ (the density $\rho_\mathrm{med}$ does not enter the equation).
Taking $\rho_\mathrm{big} \sim 0.5 \ldots 1 \g\cm^{-3}$ for large planetesimals
and $\rho = 2.35 \g\cm^{-3}$ for dust (see Appendix~\ref{ss:ACE sims}) would reduce
the total disc mass estimates by a factor of two to five.
If, instead, we require the differential {\em mass} distribution to be continuous,
Eq.~(\ref{eq:disc mass}) will remain valid as written.
We expect that the latter model, i.e., requiring that the mass distribution stays continuous at 
the density jumps, is a better approximation to reality, simply because it is the mass and not the 
number of particles that is conserved in destructive collisions.\footnote{In more detail,
\citet{wyatt-et-al-2011} showed that in steady state it is the mass loss
rate from logarithmic bins that remains the same across the whole size distribution.
This means that there would be a discontinuity in the mass distribution at such transitions,
such as that noted in \citet{wyatt-et-al-2011} at the transition from the strength to gravity regime
(see their section 2.6.1).
However, these discontinuities are expected to be small, and so not significantly affect
the calculations here.}
Thus the values of $M_\mathrm{disc}$ listed in Table~\ref{tab:sample} and
depicted in Fig.~\ref{fig:mass&xm-fd_3.5_1.0mm_3.7_1.0km_2.8_200km_slide} should hold
for the size-dependent density.

Further, if planetesimals in extrasolar systems are pebble piles or rubble piles,
they will be quite easy to destroy even in low-energy collisions.
This primarily applies to bodies with sizes between $\sim \cm$ and $\sim \km$,
i.e., between the strength and gravity regimes.
This means that more realistic simulations would have to consistently use a
``pebble-pile''-like critical fragmentation energy
prescription \citep[see Eqs.~2--4 in][and their Fig.~1]{krivov-et-al-2018}.
Applying it would result in a deep minimum in the size distribution at sizes 
between $\sim\cm$ and $\sim\km$ \citep[see Fig.~4 in][]{krivov-et-al-2018}.
However, it is possible to show that the presence of this minimum is unimportant
for the discussion of the mass budget, as long as the disc is in a collisional steady state.
This is because the amount of the observed dust is given by a product of the grain lifetime
and the dust production rate, where the latter is equal to
the mass loss rate of the largest bodies that are in the collisional equilibrium (because in
a steady-state system all the mass lost by bigger objects is transferred to smaller objects
at the same rate).
The mass loss rate of the biggest planetesimals, in turn,
is proportional to the amount of those bodies in the system.
As a consequence, the amount of dust in a steady-state disc
for a given amount of the largest planetesimals that are in the collisional equiulibrium
is independent of the fragmentation efficiency of intermediate-sized bodies.
Nevertheless, our $Q_\mathrm{D}^*$ prescription outside the fragile mid-size range
also has uncertainties of a factor of few. This may affect the values of $s_\mathrm{km}$
derived in our simulations and so the estimates of the total disc mass
(see Sect.~\ref{ss:recent ignition}).

\subsection{Dependence of disc mass on the host star and disc extent}

We now look at the total disc mass against parameters other than the
dust fractional luminosity. 
Since our sample contains only 20 discs, we cannot make any statistically significant conclusions.
Nevertheless, this analysis may show which parameters make some discs more affected
by the mass problem than the others, thus giving some hints about the origin of the problem.
Besides, we check whether any known trends are also visible in our sample. 
 
Two upper panels in Fig.~\ref{fig:mass vs various} check how the estimated mass is related to
the disc geometry.
Since the disc mass is proportional to $R^2 (\Delta R /R) \Sigma(R)$,
where the surface density of solids $\Sigma(R) \propto R^{-p}$ is commonly assumed to gently
decrease with the distance from the star ($1 \la p \la 2$), we may expect that larger discs tend to be more massive.
This is neither supported nor ruled out by Fig.~\ref{fig:mass vs various}a; the
best fit is $M_\mathrm{disc} \propto R^{0.17 \pm 0.37}$.
Similarly, we may expect radially extended discs
to have higher masses on the average compared to the narrow ones.
Figure~\ref{fig:mass vs various}b confirms this, albeit at a low significance level.
The best fit is $M_\mathrm{disc} \propto \Delta R^{0.22 \pm 0.20}$.
Note, however, that two of the three lowest-mass discs (Fomalhaut and $\varepsilon$~Eri) are the narrowest.

The remaining two panels in Fig.~\ref{fig:mass vs various}
depict the disc masses against stellar parameters.
There might be a weak trend towards higher debris disc masses
around more massive stars (Fig.~\ref{fig:mass vs various}c).
We find $M_\mathrm{disc} \propto M_*^{1.63 \pm 0.79}$.
Besides, the masses of discs around the two lowest-mass stars in the sample,
AU~Mic and $\varepsilon$~Eri, are lower that those of the other discs.
This could possibly be attributed to that fact that debris disc progenitors,
protoplanetary discs, tend to have lower masses around lower-mass primaries (see Sect.~\ref{ss:max mass}).
However, there are various observational biases here that need to be considered
given that this is not a statistical sample.
Finally, Fig.~\ref{fig:mass vs various}d  demonstrates that three discs around older stars
($0.4$--$1.4\Gyr$) in our sample reveal about an order of magnitude lower masses than
those in young systems ($10$--$40\Myr$). The formal fit through the data
yields $M_\mathrm{disc} \propto T_\mathrm{age}^{-0.70 \pm 0.34}$.
This likely reflects a well-known long-term decay of debris discs (see Sect.~\ref{ss:min mass}).

\section{Possible solutions}
\label{s:solutions}

\citet{krivov-et-al-2018} listed and briefly discussed several possible solutions
to the debris disc mass problem. Here we revisit some of them and consider
a few other possibilities.

\subsection{Uncertainty in the dust mass?}
\label{ss:opacity}

Since the total mass of a debris disc is directly proportional to the dust mass inferred
from observations (Eq.~\ref{eq:disc mass}), the question is, how reliable that dust mass is.
The largest uncertainty in calculating it arises from the mass absorption coefficient $\kappa$
in Eq.~(\ref{eq:dust mass}).
In the above analysis we set it to 
$1.7\cm^{2}\g^{-1} (850\mum/\lambda)$,
which is a common choice in the (sub)mm literature.
However, the actual opacity depends on a number of factors.
Indeed, the opacity at a given wavelength is equal to the ratio of the absorption cross section of
the disc material and the dust mass:
\be
 \kappa (\lambda)
 =
 {
   \int\limits_{s_\mathrm{min}}^{s_\mathrm{max}}
   \pi s^2 n(s) Q_\mathrm{abs}(s, \lambda) ds
 \over       
   \int\limits_{s_\mathrm{min}}^{s_\mathrm{mm}}
   \frac{4}{3}\pi \rho_\mathrm{d} s^3 n(s) ds 
 } .
\label{eq:kappa}
\ee
Note that
the sub-mm flux in a $q \approx 3.5$ size distribution is dominated by particles in the size range
from $\sim 100\mum$ to $\sim 10\cm$ \citep[see, e.g., Fig.~5 in][]{wyatt-dent-2002}.
For some materials, even grains larger than $\sim 10\cm$ can make non-negligible
contribution to sub-mm fluxes \citep{loehne-2020}.
This is why the integral in the numerator extends over the full range of sizes from
$s_\mathrm{min}$ to $s_\mathrm{max}$, thus including not only dust, but also larger material,
since it may contribute to the absorption cross section
and thus also to the observed thermal emission fluxes $F_\nu$ in Eq.~(\ref{eq:dust mass}).
In contrast, the integral in the denominator is taken from
$s_\mathrm{min}$ to $s_\mathrm{mm} = 1\mm$ to represent the mass of {\em dust}.
Defined this way, the opacity $\kappa$ gives the emitting cross-sectional area
per mass in $<1\mm$ grains assuming the size distribution to extend to larger sizes.

Dust opacity in astrophysical environments has been in the focus of interest
for decades, motivated by studies of interstellar medium in general,
star-forming clouds
and protoplanetary discs
\citep[e.g.,][and references therein]
{draine-lee-1984,beckwith-sargent-1991,pollack-et-al-1994,henning-et-al-1995,semenov-et-al-2003,draine-2006}.
In all these applications,
most of the mass is thought to be in dust rather than big objects.
Consequently, the vast majority of previous studies on opacity
considered single grain sizes
or a size distribution that is truncated at a size 
on the order of mm or cm.
In the latter case, it is a common practise to use both integrals in Eq.~(\ref{eq:kappa})
with the same upper limit that is set to somewhere in the mm to cm range.
While this is sufficient for distributions in which there is little mass in objects larger than cm,
this is not the case in a debris disc.
This is why in the following discussion it is essential to use Eq.~(\ref{eq:kappa}) as written, with
the caveat that caution is required when comparing opacities discussed here with those reported
in the literature.

Equation~(\ref{eq:kappa}) shows that the opacity depends on
the material composition and morphology of dust grains (which determine the absorption efficiency
$Q_\mathrm{abs}$ and the bulk density of dust $\rho_\mathrm{d}$).
The minimum size $s_\mathrm{min}$ is closely related to the radiation pressure
blowout limit \citep[e.g.,][]{pawellek-krivov-2015} and so, also depends on the dust composition.
However, the dependence on $s_\mathrm{min}$ is weak, since particles of size $s_\mathrm{min}$ emit
inefficiently and contain little mass.
Finally, the opacity depends on the size distribution $n(s)$~-- not only at dust sizes
but, for some materials, also at larger sizes if these contribute to the thermal emission
at wavelength $\lambda$.

To see to what extent all these parameters may alter the opacity,
we chose $\lambda = 1\mm$ as a reference wavelength, lying between the two ALMA bands
($\approx 850\mum$ and $\approx 1.3\mm$) in which discs of our sample were observed.
We start with considering ``astronomical silicate'' \citep{draine-2006}
as the dust composition, which is a traditional (although not really justified) choice in debris disc modeling. 
Its bulk density was taken to be $\rho_\mathrm{d}=3.5\g\cm^{-3}$, and the absorption efficiency
was calculated from the optical constants by using the Mie algorithm of \citet{wolf-voshchinnikov-2004}.
Assuming $q=3.5$ and $s_\mathrm{min}=1\mum$,
the resulting opacity $\kappa = 5.14 \cm^2\g^{-1}$.
However, it varies with the size distribution slope and the minimum grain size.
For $3\le q \le 4$ and $s_\mathrm{min} = 1\mum$,
we determine $2.34 \cm^2\g^{-1} \le \kappa \le 14.6 \cm^2\g^{-1}$.
As noted above, the dependence on $s_\mathrm{min}$ is weak, and is only noticeable
for steeper size distributions, for which smaller particles make a larger contribition to
the cross section.
As an example, assuming $q=4$ and varying $s_\mathrm{min}$ from $0.5\mum$ to $5\mum$
changes $\kappa$ from $2.15 \cm^2\g^{-1}$ to $2.95 \cm^2\g^{-1}$.

Since we do not know what the composition of debris dust looks like in reality,
we can invoke results of laboratory measurements with THz techniques performed for
some materials of potential astrophysical relevance.
As an example, we consider crystalline water ice \citep{reinert-et-al-2014},
forsterite \citep{mutschke-mohr-2019}, 
pyroxene (H. Mutschke \& P. Mohr, in prep.),
and amorphous carbon \citep{zubko-et-al-1996}.
At $\lambda = 1\mm$
and a temperature of $100\K$,
and assuming a size distribution with  $q=3.5$ and $s_\mathrm{min} = 1\mum$,
the opacity varies from
$\sim 0.48 \cm^{2}\g^{-1}$ for forsterite and
$\sim 1.34 \cm^{2}\g^{-1}$ for crystalline water ice
through
$\sim 3.30 \cm^{2}\g^{-1}$ for pyroxene to
$\sim 14.4 \cm^{2}\g^{-1}$ for amorphous carbon.
Interestingly, the opacity  $1.7\cm^2\g^{-1}(850\mum/\lambda)$
assumed here, or $1.45\cm^2\g^{-1}$ at $1\mm$, 
is in the middle of the range for these extreme assumptions.

By analogy with compositional constraints derived for the Solar system comets
\citep[e.g. 67/P, see][]{fulle-et-al-2016}, it is likely that the actual debris dust
is a mixture of pure materials like these.
However, neither the fractions of the components nor the way they are mixed (e.g., layer-like or
inclusion-like) are known.
Additional uncertainties come from (unknown) porosity of grains \citep{henning-stognienko-1996},
as well as from the temperature dependence of the opacity.
For all of the materials listed above, opacity tends to decrease toward lower temperatures
\citep[][H. Mutschke \& P. Mohr, in prep.]
{reinert-et-al-2014,haessner-et-al-2018,potapov-et-al-2018,mutschke-mohr-2019}.

To summarize, the assumed debris dust opacity
is uncertain, perhaps by a factor of several.
We argue, however, that this uncertainty is not of crucial importance for
the mass problem.
For example, were the opacity  of the debris material lower than commonly assumed,
correcting for this would further
increase the inferred dust mass and so also the total mass of the discs.
However, this could equally affect the inferred masses of protoplanetary discs
and thus proportionally increase the ```maximum mass of a debris disc.''
As a result, this would just ``upscale'' the entire mass problem without increasing or reducing
the tension between the estimated and maximum possible masses of debris discs,
albeit implying that PPDs are more prone to gravitational instability \citep{toomre-1964}.

\subsection{An overall steeper size distribution?}

The above discussion of the mass problem relies upon models of the collisional
cascade that predict a certain slope of the size distribution across many orders
of magnitude of particle sizes, from mm-sized grains to km-sized objects~-- i.e.,
the largest objects that can be destroyed by collisions within the age of the system.
The question arises, how certain these predictions actually are.
Should, for some reasons,
the overall size distribution slope be somewhat steeper, the same amount of dust could
be sustained by a planetesimal population of lower mass.
One possibility to change the slope is to allow the mean relative velocities $v$
in the disc to be size-dependent.
Assuming that $Q_\mathrm{D}^* \propto s^\gamma$ and $v \propto s^p$,
\citet{kobayashi-tanaka-2010} and \citet{pan-schlichting-2012} showed that
the slope of the differential size distribution is given by
\be
  q = {21 + \gamma - 2p
      \over 
        6 + \gamma - 2p},
  \label{eq:q(gamma,p)}
\ee 
leading for $p > 0$ to steeper slopes than those computed under the assumption
of size-independent relative velocities.

Several effects are responsible for the evolution of the relative velocities amongst particles
in a debris disc \citep[e.g.,][]{stewart-wetherill-1988}.
In addition to gas which is discussed in Sect.~\ref{ss:gas},
the effects of viscous stirring, damping by inelastic collisions, and damping by dynamical friction
are most important and discussed in Sect.~\ref{ss:damping}.

\subsubsection{Collisional damping}
\label{ss:damping}
\citet{pan-schlichting-2012} found analytically
power-law solutions for the combined steady-state
size and velocity distributions in collisional debris discs.
In one of the regimes they considered (see their Fig.~3), they found $p$
to lie between $0.16$ and $0.5$ in the size range from $1\m$ to $\gg 1\km$,
and confirmed this analytic prediction by direct numerical simulations
with a collisional code. There are no reasons to expect that this moderately positive
slope $p$ would be different below the lowest cutoff of their setup, $1\m$,
down to the smallest objects still unaffected by radiation pressure,
say $\sim 1\cm$.

Based on this result, let us now conservatively assume $p=0.2$ in the size interval
from $\sim 1\cm$ to $\sim 1\km$. Taking $\gamma = -0.34$ as appropriate for
the strength regime, Eq.~(\ref{eq:q(gamma,p)}) gives $q=3.85$. Plugging this into our
toy model including the primordial size distribution from planetesimal formation
simulations, and applying to the same sample of debris discs as before,
we estimated disc mass would reduce by a factor of~8.
The mass problem would essentially be solved, at least for the majority of the discs
in our sample.

If this effect is at work, this might be in line with the in-depth
collisional simulations that aimed at reproducing the observed resolved
thermal emission and/or scattered light images of several individual systems
\citep[e.g.,][]{reidemeister-et-al-2011,loehne-et-al-2011,schueppler-et-al-2014,%
schueppler-et-al-2015,schueppler-et-al-2016},
from which parent planetesimals with eccentricities
$\sim 0.01$...$0.05$ were inferred.
These values are lower than those typical for the Kuiper belt, and
also lower than those that might be expected from the presence
of known stirring planets in some systems (such as HR 8799).
However, the conclusions that the eccentricities are low were based on ACE simulations
that assumed the same excitation across the range of sizes, which is a limitation
of the current ACE code. Strictly speaking, these conclusions apply to
the eccentricities of cm-sized grains (the destruction of which produces
visible dust), and not to those of larger planetesimals.
Had collisional damping been included, and had it been efficient, these lower eccentricites
of cm-sized grains could have resulted from a more stirred population of
km-sized and larger planetesimals.
In the above example where $p=0.2$,
the latter would be an order of magnitude higher.

However, the work by \citet{pan-schlichting-2012} rests on a number of
assumptions that all were necessary to keep the problem
treatable analytically. 
One of them is that bullet-target size (or mass) ratios in destructive collisions are
close to unity.
However, for high-mass-ratio collisions, the damping rate is reduced by
a factor of $(m_\mathrm{p}/m_\mathrm{t})^2$,
where $m_\mathrm{p}$ and $m_\mathrm{t}$ are masses of colliders
($m_\mathrm{p} \ll m_\mathrm{t}$), see Eq. (17) in \citet{kobayashi-et-al-2016}.
This is actually what one would expect under typical debris disc conditions.
Indeed, a few debris disc simulations done so far with collisional codes that
include viscous stirring and collisional damping effects, do not
show any noticeable eccentricity damping towards smaller objects
in the centimetre-kilometre size range~--
see, e.g., Fig.~1 right in \citet{kenyon-bromley-2008}
or Fig.~3 in \citet{kobayashi-loehne-2014}.

One more uncertainty is related to the point made in Sect.~\ref{ss:formation} about planetesimals possibly being
``pebble piles'', whose critical fragmentation energy $Q_\mathrm{D}^*$ in the
strength regime is very different from the one of the ``monolithic'' bodies
\citep{krivov-et-al-2018}.
For such $Q_\mathrm{D}^*$, the analytic model by \citet{pan-schlichting-2012}
is no longer valid. To our knowledge, the role of the stirring-damping
balance in such systems has never been tested in numerical simulations either.
We expect that this would make damping less relevant, as the planetesimals are weaker and
so the ratio $m_\mathrm{p}/m_\mathrm{t}$ is even smaller than in the ``monolithic case.''
However, should a regime in which collisional damping is strong be found
in the future, this could provide a natural solution
to the disc mass problem.

\subsubsection{Secondary gas}
\label{ss:gas}

Velocity damping may also come from the ambient gas, if there is some in the disc.
Debris discs are generally known to be gas-poor, yet a handful of young 
debris discs, mostly around earlier-type stars, reveal gas in detectable amounts 
\citep[e.g.,][]{kral-et-al-2017c}.
This gas, notably CO, is likely secondary and is released from planetesimals in 
collisions, i.e., in the same process that creates observable debris dust.
Figure~\ref{fig:mass vs various}d shows that in many of the discs with high masses
gas has indeed been detected.

To check to what extent gas damping can alter the size distribution,
we consider the same system as in runs~A and B of Appendix~\ref{ss:ACE sims}:
an A-type star ($2.16M_\odot$, $27.7 L_\odot$) as a primary
and a disc with a radius of $100\AU$ and width of $10\AU$.
However, we now take an extreme case of the disc
containing $\sim 0.06 M_\oplus$ of CO gas
\citep[as appropriate for the most gas-rich debris discs such as 
HD~21997 or HD~131835,][]{kral-et-al-2017c}.
The gas is assumed to be co-located with the dust.

Assuming subsonic regime, the stopping time 
\citep[see, e.g.,][]{weidenschilling-1977}
is given by
$
 T_\mathrm{stop}
 =
 (\rho_\mathrm{g} / \rho_\mathrm{d})
 (s / c_\mathrm{s})
$,
where
$\rho_\mathrm{g}$ is the spatial density of gas,
$\rho_\mathrm{d}$ the bulk density of dust
(assumed to be $2.35 \g\cm^{-3}$),
and
$c_\mathrm{s} = \sqrt{(k T_\mathrm{g}) / (\mu m_\mathrm{H})}$
the sound speed in the ambient gas.
In the last equation, $k$ is the Boltzmann constant, $\mu$ the molecular weight
(which we set to 28 as appropriate for CO gas), and $T_\mathrm{g}$ the gas
temperature which, for the sake of rough estimates, is taken to be the
black body temperature at the disc radius:
$T_\mathrm{g} \approx 278\K (L_*/L_\odot)^{1/4} (R/\AU)^{-1/2}$.
For the gas-rich disc around an A-type star considered here,
the stopping time is
\be
 T_\mathrm{stop}
 \approx
 330 \left(s \over 1\mum\right) \yr .
\label{eq:T_stop}
\ee

The stopping time is also the timescale on which orbital eccentricities and inclinations
of particles can be substantially damped by gas.
However, efficient damping is only possible if the damping timescale is shorter that
the collisional timescale of the disc solids.
The latter can be estimated as \citep[e.g.,][]{wyatt-2005}
\be
 T_\mathrm{coll}
 \approx
 {P \over 8 \pi f_\mathrm{d}}
 {\Delta R \over R} ,
\label{eq:T_coll}
\ee
where $P = \sqrt{G M_* / R^3}$ is the orbital period.
This is the collisional lifetime of the smallest grains that dominate the cross section
(or the optical depth) of the disc. We assume these to have size just above the radiation
pressure blowout limit which, for the A-star and grain density of assumed here (see
Appendix~\ref{ss:ACE sims}), is equal to $6.3 \mum$.
For a size distribution with a slope $q$, the collisional lifetime of larger grains
scales as $ T_\mathrm{coll}(s) \propto s^{4 - q}$
\citep[e.g.,][]{wyatt-dent-2002,o'brien-greenberg-2003,wyatt-et-al-2011}.
For the same disc and the same A-type star as above,
and assuming the $f_\mathrm{d} = 10^{-4}$,
we obtain
\be
 T_\mathrm{coll}
 \approx
 1.1 \times 10^4 \left(s \over 1\mum \right)^{4-q} \yr .
\label{eq:T_coll_num}
\ee
From Eqs.~(\ref{eq:T_stop}) and~(\ref{eq:T_coll_num}) with $q=3.5$,
we find the condition for efficient damping,
$T_\mathrm{stop} < T_\mathrm{coll}$, to be fulfilled for all grains smaller than about $1\mm$ in size.
As a result, we expect steepening of the size distribution at those sizes
(a positive $p$ in Eq.~\ref{eq:q(gamma,p)}).
Said differently, this should result in an enhanced dust density compared to that of larger
solids. 

Note that the total amount of  gas mass in such discs can be even higher, as most of the
gas mass could be in atomic C and O. In some of the discs, this has directly been confirmed
observationally. As an example, in the HD 32297 disc
a CO mass of $4\times 10^{-4} M_\oplus$ \citep{macgregor-et-al-2018b}
and a neutral C mass of $3.5 \times 10^{-3} M_\oplus$ \citep{cataldi-et-al-2020}
have been inferred.
A larger total mass of gas would mean that gas damping can be efficient at sizes larger
than inferred from the above calculation, leading to a stronger alteration of the size distribution compared to
the gas-free case.
However, in extreme cases, such as the discs with 
$\sim 0.06 M_\oplus$ of CO gas quoted above, CO might be self-shielded and so the mass in CO
could be comparable to or even greater than that of C
\citep[see for example Fig.~6 of][]{marino-et-al-2020b} and so the above calculation would be unaffected.

Simple estimates of how strong the effects would be are difficult to make
for several reasons. For instance, sufficiently small grains can be swiftly
damped so strongly that collisions with like-sized grains would no longer be disruptive.
On the other hand, these grains would also collide with larger ones that retain
higher dynamical excitation.
Sufficiently accurate quantitative predictions for these effects would require
a dedicated modeling, taking into account a size distribution and a size-dependent
distribution of eccentricities and excitations of grains
in the ambient gas environment.

\subsubsection{Any signs of damping in mm observations?}
\label{ss:damping at mm}

In the previous subsections, we argued that  damping is not likely to be efficient,
except perhaps for sub-mm sizes in some discs with appreciable amount of gas.
Regardless of theoretical arguments, it is useful to see whether any signs of damping
at such sizes can be seen in observations.
Debris disc observations at multiple long (mm to cm) wavelengths to get
the spectral index of the emission can be used to constrain size distributions
of grains in the size range comparable to the wavelength.
A standard method is to use the \citet{draine-2006} equation that relates
the measured SED slope in the mm regime 
to the slope of the dust size distribution.
Applying this method to five discs observed with ATCA at $9\mm$,
\citet{ricci-et-al-2015b} inferred
$3.36 < q < 3.50$, with a typical uncertainty of $0.07$--$0.08$.
A similar analysis for a larger selection of discs observed with several facilities
at wavelengths from
$0.85\mm$ to $9\mm$ was done by \citet{macgregor-et-al-2016} who derived
$3.2 < q < 3.7$, with a weighted mean of $q=3.36 \pm 0.02$.
Most recently, \citet{loehne-2020} extended the \citet{draine-2006} formula
by lifting several assumptions that 
are not necessarily fulfilled for debris discs.
Applied to a sample similar to that of \citet{macgregor-et-al-2016},
this model yields the mean slope of $q = 3.42 \pm 0.04$ and $q = 3.55 \pm 0.05$,
assuming dust composition to be astrosilicate and pyroxene, respectively.

The disc samples used in these analyses included some of the discs that are also part of the 
\citet{matra-et-al-2018} sample used in this paper.
Some of these (e.g., HD~95086 and HD~131835) are 
among those strongly affected by the mass problem,
yet their inferred size distribution slopes are rather low ($q<3.5$).
This might argue against steeper distribution, and so also against efficient damping 
in such discs, at least at sizes smaller than $\sim \cm$.
Of course, for damping to work, the size distribution only needs to be steep at cm--km sizes.
However, the damping (be that by gas or collisions) is expected to be more efficient
at smaller not larger sizes (e.g., the size of bullets goes down as $Q_\mathrm{D}^*$
increases to smaller sizes, and gas drag gets stronger).
Since there is no evidence for it at smaller sizes, it is then unlikely that it is at work
at larger ones.

\subsection{Is the ``collisional age'' important?}
\label{ss:recent ignition}

It is possible that for some reason the collisional cascade in some systems started, or at least
became intensive, very recently and not right after the completion of the protoplanetary phase.
This can be caused, for instance, by late planetary instabilities
\citep[e.g.,][]{booth-et-al-2009,raymond-et-al-2011,raymond-et-al-2012}
or by recent stellar flybys 
\citep[e.g.,][]{kobayashi-ida-2001,kenyon-bromley-2002}.
In that case, a debris disc even around an old, $\sim\Gyr$-aged, star could be ``collisionally young.''

To see how the ``collisional age'' of a disc may affect the disc mass estimates, consider
Eq.~(\ref{eq:disc mass}).
The only quantity in Eq.~(\ref{eq:disc mass}) that depends
on system's ``collisional age'' is $s_\mathrm{km}$
(see discussion in Sect.~\ref{ss:formation}).
To get quantitative estimates for this dependence,
we use again the two collisional simulations with the ACE code
(runs~A and~B) described in Appendix~\ref{ss:ACE sims}.
Remember that the largest planetesimals were $200\km$ in radius in both cases, 
and the only difference between the two runs was the assumed initial size distribution slope of
the primordial population of objects with radii from $1\km$ to $200\km$.
In run~A, we set it to $3.7$, i.e., made it the same as for sub-km planetesimals.
In run~B, we took $2.8$, as predicted by pebble concentration models of planetesimal formation.

\begin{figure}
\hspace*{-5mm}
\includegraphics[width=1.12\columnwidth]{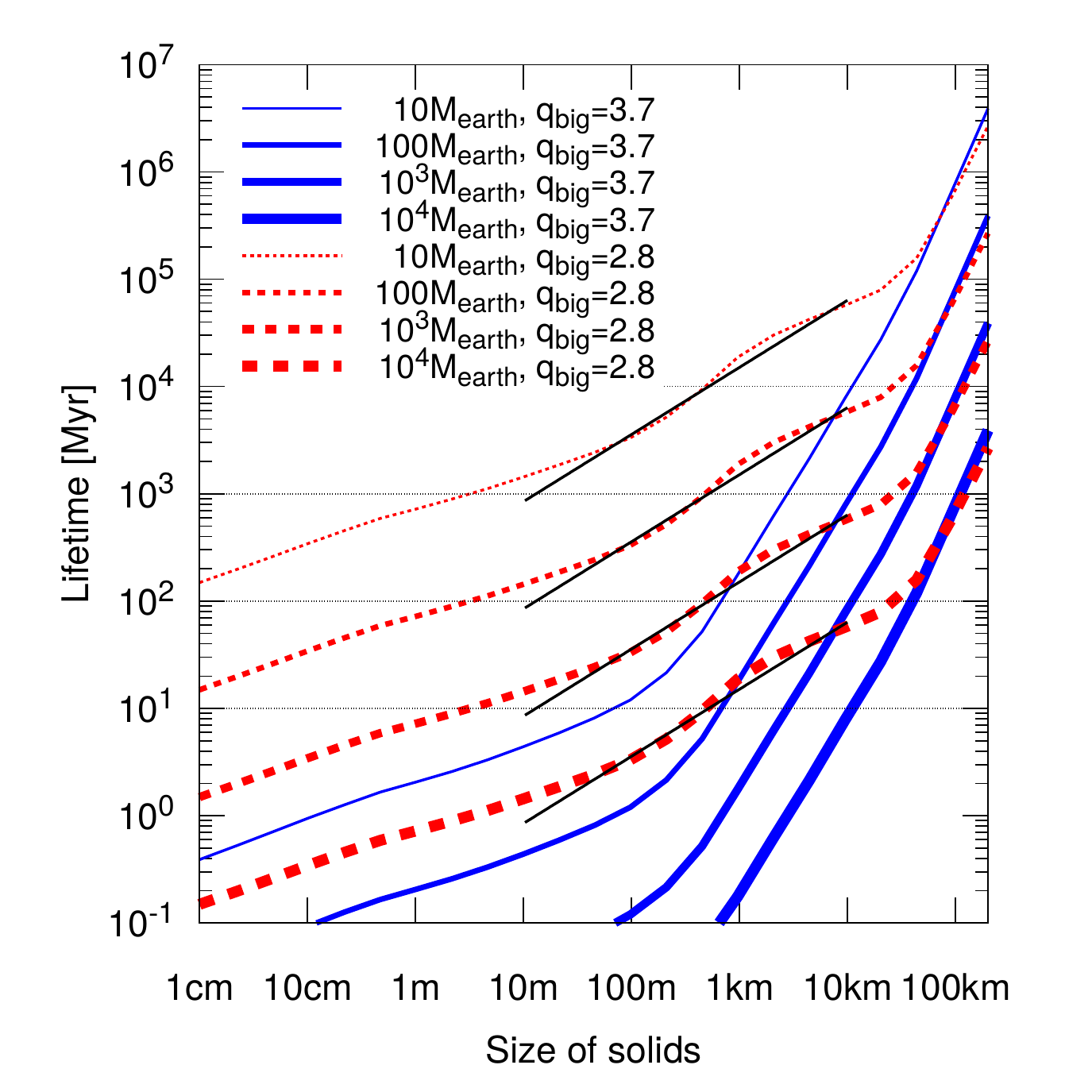}
    \caption{Collisional lifetime of bodies in two groups of fiducial debris discs
    (runs~A and~B in Appendix~\ref{ss:ACE sims}).
    These groups assume different size distribution slopes for a primordial population of planetesimals:
    $q_\mathrm{big} = 3.7$ (run~A, solid blue lines) and
    $q_\mathrm{big} = 2.8$ (run~B, dashed red ones).
    Each group contains four discs with different total masses of
    $10 M_\oplus$,
    $100 M_\oplus$,
    $10^3 M_\oplus$, and
    $10^4 M_\oplus$
    (lines of increasing thickness).
    See Appendix~\ref{ss:ACE sims} for other simulation parameters.
    Tilted black lines are power-law fits in the $10\m$ to $10\km$ range,
    Eq.~(\ref{eq:Tage}).
    }
    \label{fig:Tcoll}
\end{figure}

The collisional lifetimes of different-sized bodies computed in both runs (at the start of the
simulations) are shown in Fig.~\ref{fig:Tcoll}.
Although the total disc mass was set to the same value of $100 M_\oplus$ in the actual ACE simulations,
we also added curves for several other total masses from $10 M_\oplus$ to $10^4 M_\oplus$.
This was done by using a simple fact that, for discs of different mass but with the same shape of
size distributions, a collisional timescale is inversely proportional to the disc mass.

For $q_\mathrm{big} =3.7$ (run~A), the results shown
with solid lines in Fig.~\ref{fig:Tcoll}
agree with previous work
(cf., e.g., 
Fig.~7 in \citeauthor{wyatt-dent-2002} \citeyear{wyatt-dent-2002},
Fig.~20 in \citeauthor{loehne-et-al-2011} \citeyear{loehne-et-al-2011}
or Fig.~7 in \citeauthor{krivov-et-al-2018} \citeyear{krivov-et-al-2018}).
They demonstrate that the size of bodies that start to be depleted after $\sim 1\Gyr$
of evolution varies from
several km in discs with a total mass of $10M_\oplus$ to
about $100\km$ in discs as massive as $10^4M_\oplus$.
A caveat, however, is that accurate constraints on the size are difficult to get, as these depend
very sensitively on the assumptions.
While in Sect.~\ref{ss:formation} we showed that the critical fragmentation energy
$Q_\mathrm{D}^*$ in the mid-size range is unimportant, 
outside that range it does matter.
It is possible to get a much larger transition size if the km-sized and larger bodies
are assumed to be particularly weak~--- or to collide at higher velocities.
For instance, for the same Fomalhaut disc \citet{wyatt-dent-2002} and \citet{quillen-et-al-2007}
derive $\sim 1\km$ and $\sim 100\km$, respectively, which traces back to different assumptions made about the
strength of bodies and the different approximations employed in their collisional models.

For $q_\mathrm{big} =2.8$ (run~B, dashed lines in Fig.~\ref{fig:Tcoll}),
the resulting curves at sizes smaller than several tens of km
are relatively flat.
This means that $s_\mathrm{km}$ does depend sensitively
on the system's age $T_\mathrm{age}$.
In the size range from $10\m$ to $10\km$, fitting a power law to this dependence gives
\bea
 {T_\mathrm{age} \over 1\Gyr}
 &=&
 4.17
 \left( s_\mathrm{km} \over 1 \km \right)^{0.625}
 \left(M_\mathrm{disc} \over 100 M_\oplus \right)^{-1}
\nonumber\\
 &\times&
 \left(M_\star \over M_\odot \right)^{-1.33}
 \left(R \over 100 \AU \right)^{4.33} ,
\label{eq:Tage}
\eea
where, apart from the $M_\mathrm{disc}$ scaling explained above,
we have also included a $\propto R^{13/3}$ radius dependence
and a $\propto M_\star^{-4/3}$ dependence of the collisional timescale
\citep{wyatt-et-al-2007,loehne-et-al-2007}.
This gives
\bea
 {s_\mathrm{km} \over 1 \km}
 &=&
 0.102
 \left(T_\mathrm{age} \over 1\Gyr \right)^{1.60}
 \left(M_\mathrm{disc} \over 100 M_\oplus \right)^{1.60}
\nonumber\\
 &\times&
 \left(M_\star \over M_\odot \right)^{2.13}
 \left(R \over 100 \AU \right)^{-6.93} .
\label{eq:skm}
\eea

\begin{figure}
\includegraphics[width=1.03\columnwidth]{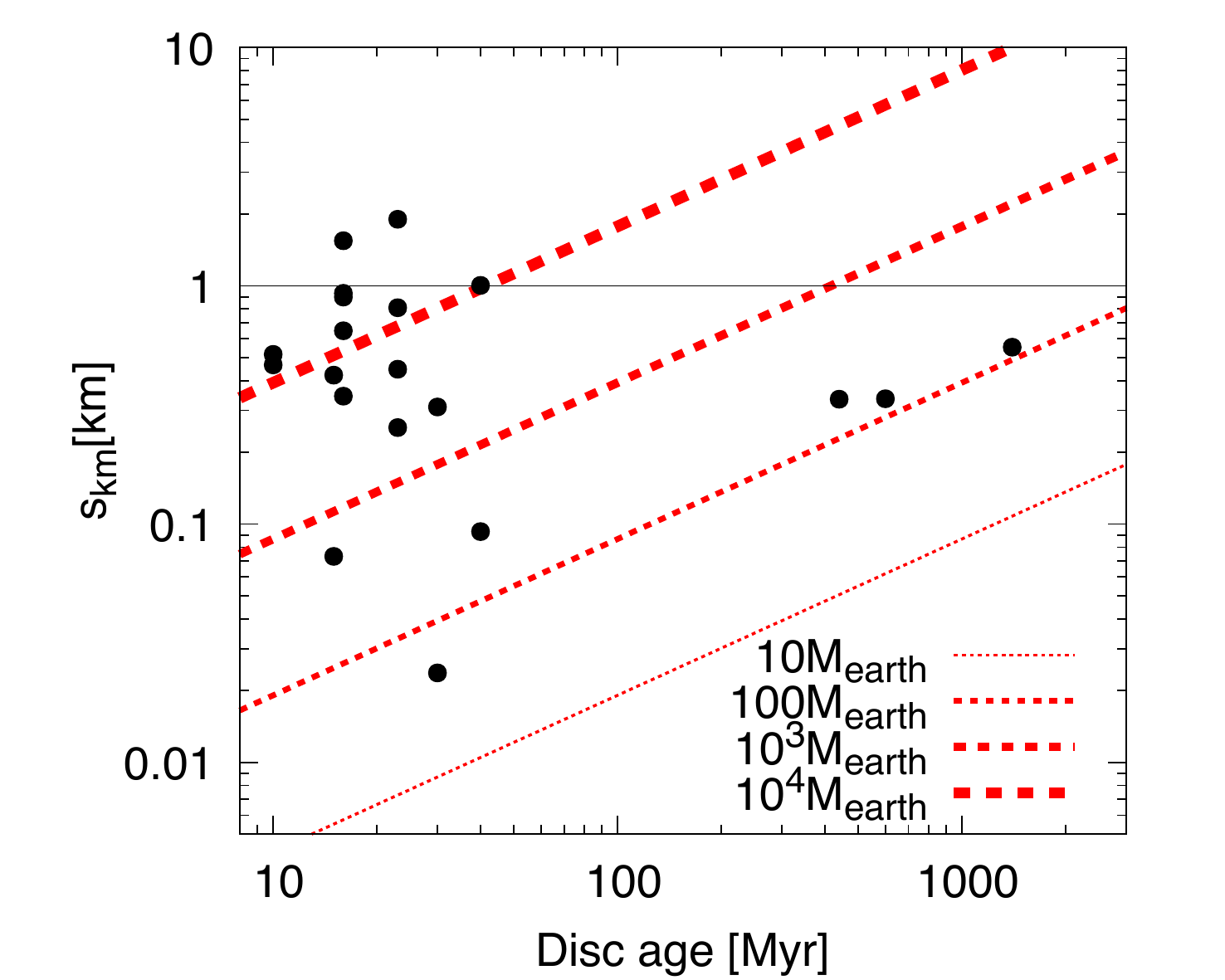}\\
\includegraphics[width=1.03\columnwidth]{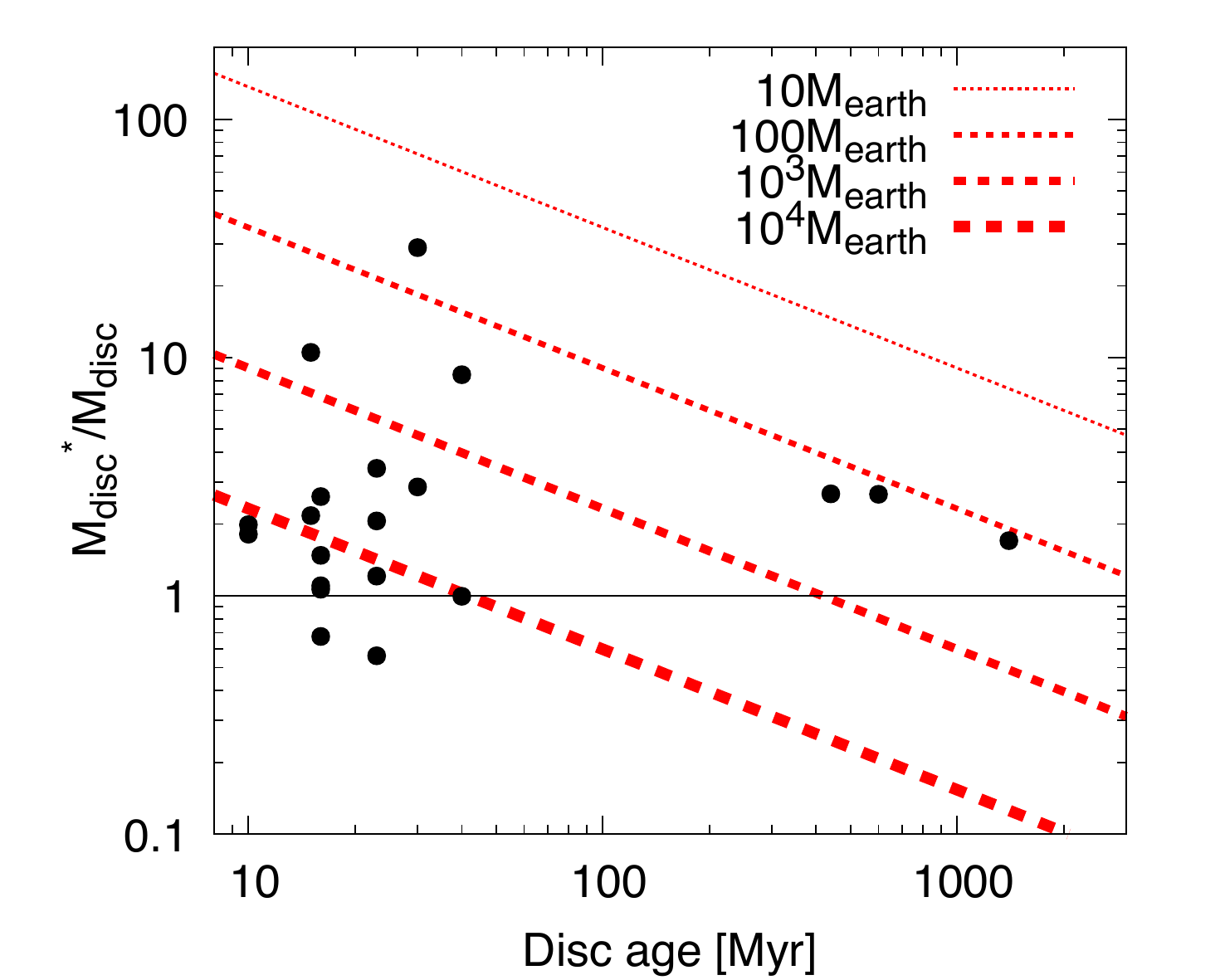}
    \caption{$s_\mathrm{km}$ (top)  and $M_\mathrm{disc}^*/M_\mathrm{disc}$ (bottom)
    as a function of disc's age.
    Lines of increasing thickness correspond to
    different $M_\mathrm{disc}$:
    $10 M_\oplus$,
    $100 M_\oplus$,
    $10^3 M_\oplus$, and
    $10^4 M_\oplus$.
    Points represent the 20 discs of our sample (without the Kuiper belt).
    }
    \label{fig:skm_and_Mdisc}
\end{figure}

We now denote by $M_\mathrm{disc}$ the disc mass calculated from
Eq.~(\ref{eq:disc mass})
by assuming $s_\mathrm{km} = 1\km$, and
by $M_\mathrm{disc}^*$ the disc mass that we would get from the same equation
if we took the ``correct'' $s_\mathrm{km}$ from Eq.~(\ref{eq:skm}).
From Eq.~(\ref{eq:disc mass}), we obtain
\be
  {M_\mathrm{disc}^* \over  M_\mathrm{disc}}
  = \left(s_\mathrm{km} \over 1 \km \right)^{q_\mathrm{big} - q_\mathrm{med}}
  = \left(s_\mathrm{km} \over 1 \km \right)^{-0.9} .
\label{eq:disc mass ratio simple}
\ee
Substituting here Eq.~(\ref{eq:skm}) (where
$M_\mathrm{disc}$ is replaced by $M_\mathrm{disc}^*$)
results in
\bea
  {M_\mathrm{disc}^* \over  M_\mathrm{disc}}
  &=&
 7.81
 \left(T_\mathrm{age} \over 1\Gyr \right)^{-1.44}
 \left(M_\mathrm{disc}^* \over 100 M_\oplus \right)^{-1.44}
\nonumber\\
 &\times&
 \left(M_\star \over M_\odot \right)^{-1.92}
 \left(R \over 100 \AU \right)^{-6.24} .
\label{eq:disc mass ratio}
\eea
Solving this for $M_\mathrm{disc}^*$ gives
\bea
 {M_\mathrm{disc}^* \over 100 M_\oplus}
 &=&
 2.32
 \left(T_\mathrm{age} \over 1\Gyr \right)^{-0.59}
 \left(M_\mathrm{disc} \over 100 M_\oplus \right)^{0.41}
\nonumber\\
 &\times&
 \left(M_\star \over M_\odot \right)^{-0.79}
 \left(R \over 100 \AU \right)^{2.56} .
\label{eq:corrected disc mass}
\eea
For any disc with a mass $M_\mathrm{disc}$ (estimated with $s_\mathrm{km} = 1\km$), 
having a certain age $T_\mathrm{age}$,
Eq.~(\ref{eq:corrected disc mass}) allows us to find the corrected mass
$M_\mathrm{disc}^*$, and
from Eq.~(\ref{eq:skm}) (using 
$M_\mathrm{disc}^*$ instead of $M_\mathrm{disc}$ there)
we then find the actual transition size $s_\mathrm{km}$.

Lines in Fig.~\ref{fig:skm_and_Mdisc} show the results as a function of the disc age
for different $M_\mathrm{disc}$.
Overplotted with points are the individual discs in our sample. The
figure demonstrates that for nearly all discs in our sample,
$s_\mathrm{km}$ is within a factor of several of the reference value, $1\km$,
and $M_\mathrm{disc}^*/M_\mathrm{disc}$ is close to unity, within a factor of a few as well.
The largest deviation of the mass ratio from unity, by a factor of~30,
occurs for the young and large disc of HR~8799.
The same is seen from Table~\ref{tab:sample} where both $M_\mathrm{disc}$
and corrected disc masses $M_\mathrm{disc}^*$ computed with Eq.~(\ref{eq:corrected disc mass})
are given. 

This means, in particular,
that assuming a younger ``collisional age'' for a certain disc would not lead to a much lower
disc mass. This is easy to understand from the above equations that are ``self-regulating.''
A younger collisional age means, according to Eq.~(\ref{eq:skm}), a smaller
$s_\mathrm{km}$.
A smaller $s_\mathrm{km}$ would imply a larger disc mass (see Eq.~\ref{eq:disc mass ratio simple}),
but a larger mass would make $s_\mathrm{km}$ larger.
This also explains why the scatter of $s_\mathrm{km}$ and
$M_\mathrm{disc}^*/M_\mathrm{disc}$ is moderate, although parameters of individual discs
vary significantly (e.g., a factor of 3 difference in radius)
and the dependence on some of them is strong
(such as $s_\mathrm{km} \propto R^{-6.93}$ as seen from Eq.~\ref{eq:skm}).

\begin{figure}
\includegraphics[width=\columnwidth]{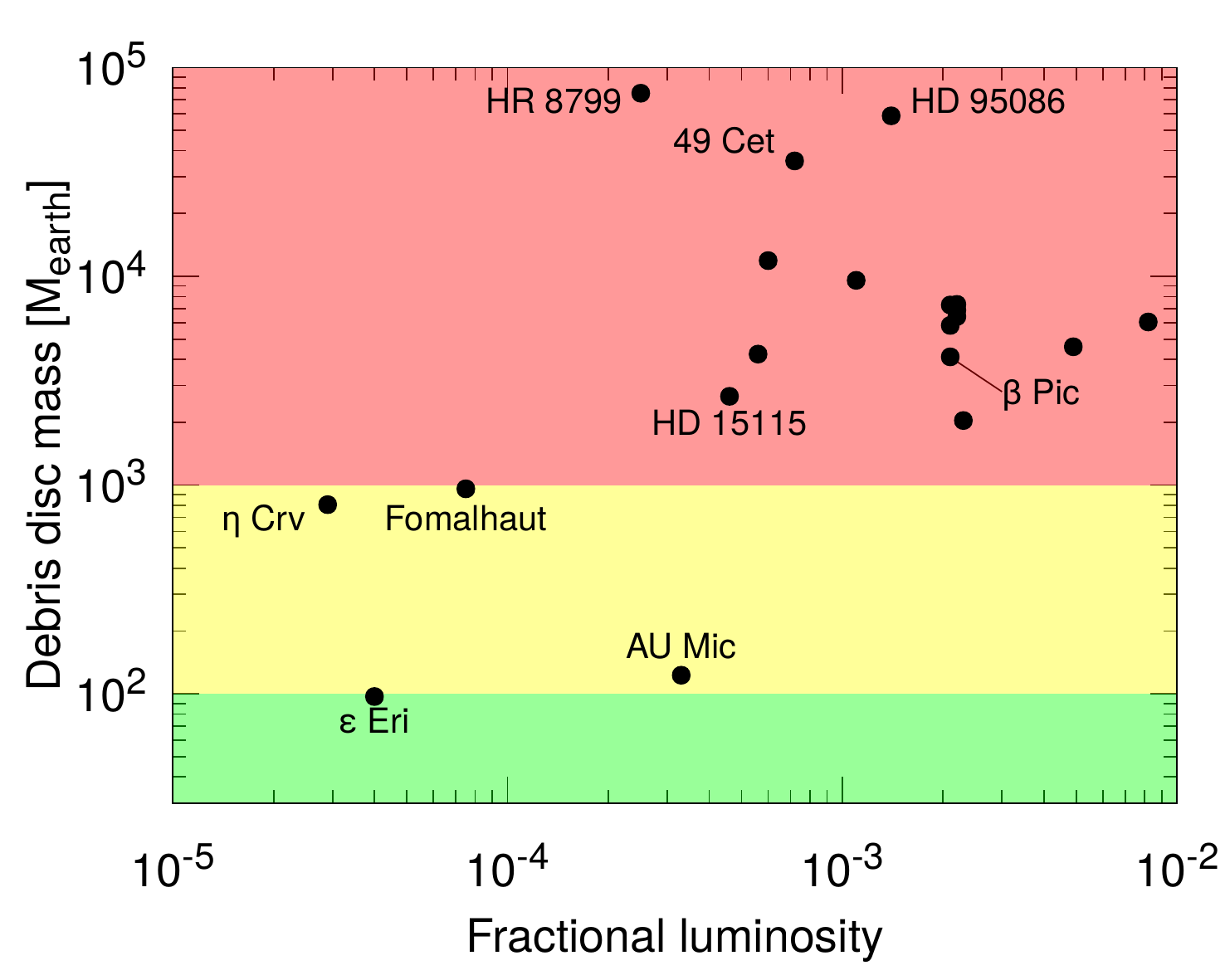}
    \caption{
      Similar to Fig.~\ref{fig:mass&xm-fd_3.5_1.0mm_3.7_1.0km_2.8_200km_slide},
      but presenting the disc mass $M_\mathrm{disc}^*$, i.e., the total disc mass 
      corrected for the time evolution of the transition size $s_\mathrm{km}$
      (Eq.~\ref{eq:corrected disc mass final}).
      Note that the plotting range is different from previous figures.
      Kuiper belt is not included.
    }
    \label{fig:corrected_mass-fd_3.5_1.0mm_3.7_1.0km_2.8_200km_slide}
\end{figure}

Inserting $M_\mathrm{disc}$ from Eq.~(\ref{eq:disc mass})
with $s_\mathrm{mm} = 1\mm$ and $s_\mathrm{km} = 1\km$
into Eq.~(\ref{eq:corrected disc mass}) yields
\bea
 {M_\mathrm{disc}^* \over 100 M_\oplus}
 &=&
 18.2
 \left(T_\mathrm{age} \over 1\Gyr \right)^{-0.59}
 \left(M_\mathrm{d} \over M_\oplus \right)^{0.41}
\nonumber\\
 &\times&
 \left(M_\star \over M_\odot \right)^{-0.79}
 \left(R \over 100 \AU \right)^{2.56}
\nonumber\\
 &\times&
 \left(s_\mathrm{max} \over 200\km \right)^{0.49} .
\label{eq:corrected disc mass final}
\eea
This equation is our final estimate of the total mass
obtained by extrapolating the dust masses to objects of size $s_\mathrm{max}$.
Figure~\ref{fig:corrected_mass-fd_3.5_1.0mm_3.7_1.0km_2.8_200km_slide} plots the
total disc mass $M_\mathrm{disc}^*$ for $s_\mathrm{max}=200\km$.
The majority of discs have masses between $10^3 M_\oplus$ and $10^4 M_\oplus$, and three discs 
(49~Cet, HD 95086 and HR 8799) are even more massive than $3\times 10^4 M_\oplus$, or $0.1 M_\odot$,
clearly exceeding the maximum possible mass of a debris disc (Sect.~\ref{ss:max mass}).

\subsection{Recent giant impacts?}

Another way the disc could appear young is if the debris was created
in the recent collisional disruption of an embryo-sized embedded body.
It is indeed possible that such events are
responsible for prominent asymmetries seen at large separations in some debris discs, such as
that of $\beta$~Pictoris \citep{jackson-et-al-2014}.
However, it is extremely unlikely that bright discs themselves are produced by giant
impacts.
There are two arguments that speak against this.
Firstly, the resulting dust disc, as long as it is bright enough, will be asymmetric.
That asymmetry is associated with the collision point and lasts for around $1000$ orbital periods,
i.e., a few Myr at about $100\AU$.
This is incompatible with a smooth, featureless spatial distribution of dust 
typically seen in (sub)mm observations of the bright discs.
And conversely, at later times when the debris ring left after a giant impact becomes symmetric,
it is no longer bright.

\begin{figure}
\includegraphics[width=1.05\columnwidth]{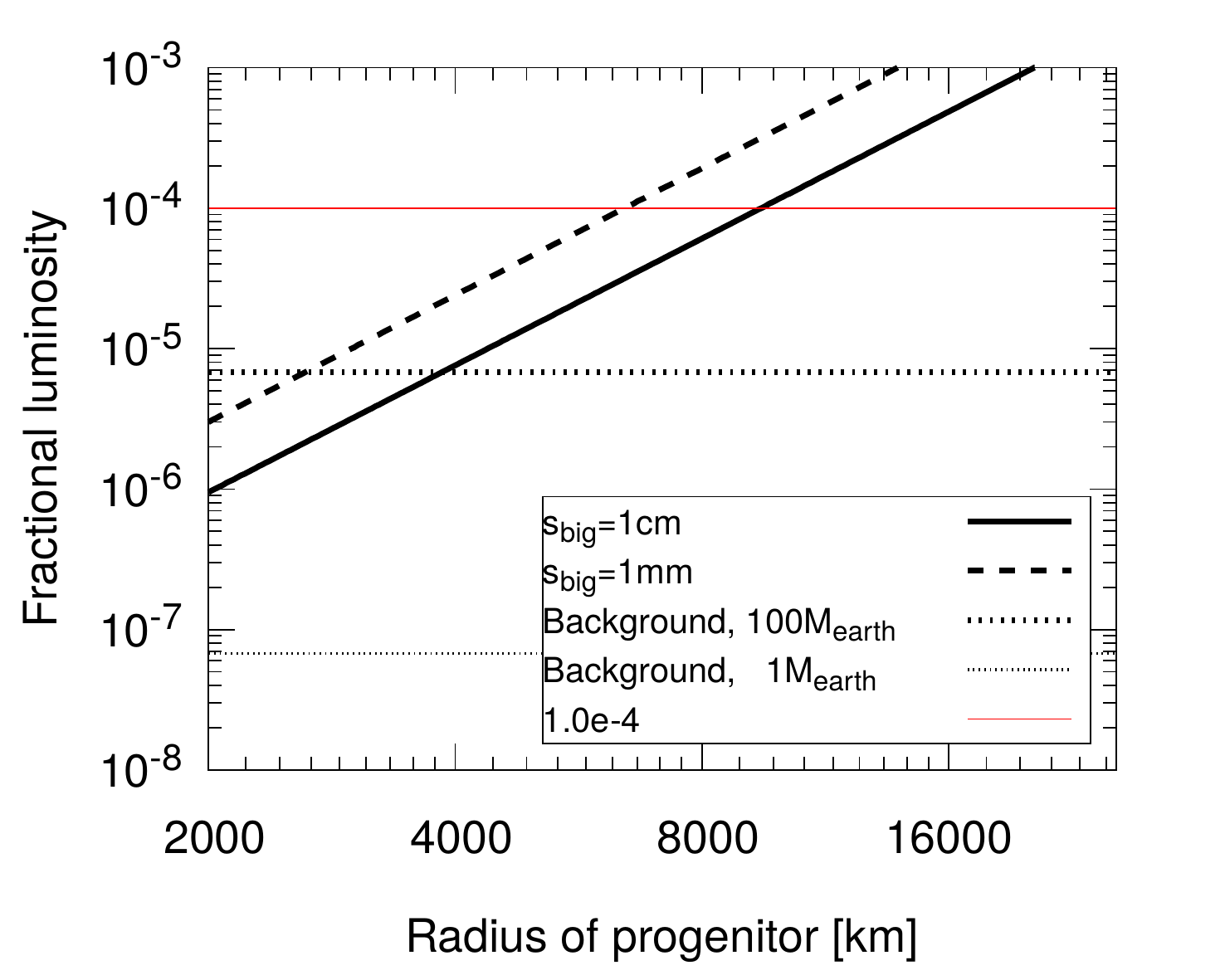}\\
\includegraphics[width=1.05\columnwidth]{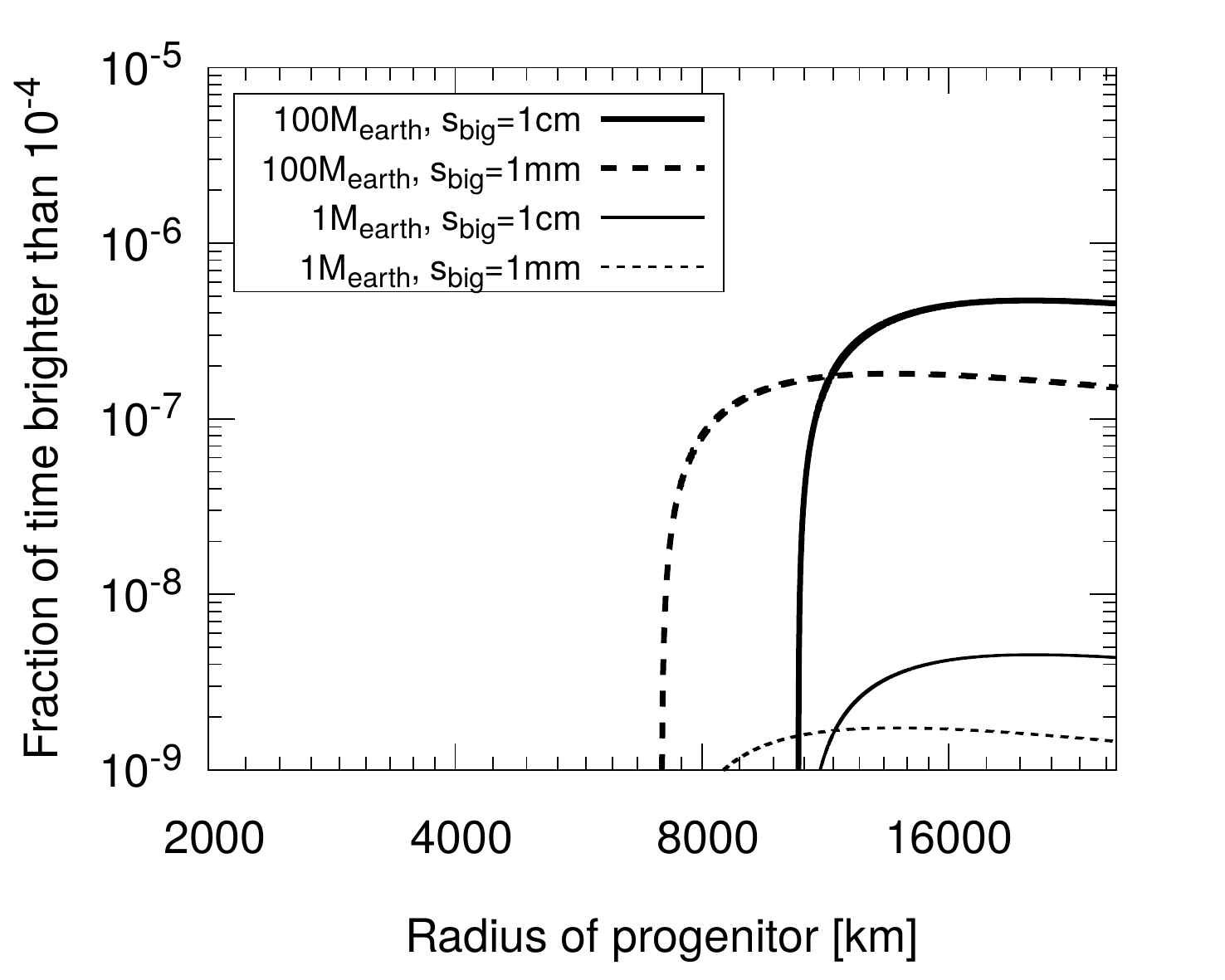}
    \caption{Brightening of debris discs by giant impacts.
     Top: peak fractional luminosity of the impact-generated debris cloud (slanted lines)
     and of the background ``regular'' disc (black horizontal lines).
     Bottom: fraction of time during which the impact debris has a fractional luminosity
     greater than $10^{-4}$.
     Thick lines: $M_\mathrm{disc} = 100 M_\oplus$,
     thin: $M_\mathrm{disc} = 1 M_\oplus$.
     Solid lines: $s_\mathrm{big} = 10\cm$,
     dashed: $s_\mathrm{big} = 1\mm$.
    }
    \label{fig:giant_impact}
\end{figure}

Secondly, it is easy to show the frequency of such events is too low.
We consider an underlying debris disc of 
mass $M_\mathrm{disc}$ in particles with radii
[$s_\mathrm{blow}$, $s_\mathrm{max}$] (with $s_\mathrm{blow}$ being the blowout size),
in which a big object (``progenitor'')
of radius $s_\mathrm{p}$ (mass $M_\mathrm{p}$) is collisionally disrupted,
releasing a fraction $\alpha$ of its mass into grains with radii
[$s_\mathrm{blow}$, $s_\mathrm{big}$].
We assume, for simplicity, the size distribution of particles both in the background disc
and the impact-generated debris cloud
to follow the Dohnanyi law with $q=3.5$.
Both the fractional luminosity of the background disc
$f_\mathrm{bkg}$
and that of the debris cloud
$f_\mathrm{cl}$
are calculated from Eq.~(15) of \citet{wyatt-2008} showing,
in particular, that
\be
f_\mathrm{cl} \propto M_\mathrm{p} s_\mathrm{big}^{-0.5}.
\label{eq:fcl}
\ee
The ratio of the two luminosities is given by
\be
{f_\mathrm{cl} \over f_\mathrm{bkg}}
=
{\alpha M_\mathrm{p} \over M_\mathrm{disc}}
\left( s_\mathrm{max} \over s_\mathrm{big} \right)^{0.5} ,
\label{eq:frac lum ratio}
\ee
and so the collisional debris can only be brighter than the background disc
if a single progenitor holds a significant fraction of the total mass,
and the collisional debris is put into small particles.

For the same set of stellar and disc parameters as in Appendix~\ref{ss:ACE sims}
and assuming $\alpha = 0.03$ \citep[as in][]{jackson-et-al-2014},
the fractional luminosities are depicted in Fig.~\ref{fig:giant_impact} (top)
for different values of $M_\mathrm{disc}$ and $s_\mathrm{big}$.
The figure shows that the debris cloud can only have fractional luminosity
in excess of $f_\mathrm{lim} \equiv 10^{-4}$ (shown with thin red line)
if the progenitor is larger than $\sim 7000$--$10000\km$
in size, depending on the $\s_\mathrm{big}$ assumed.
This threshold value of $10^{-4}$ was chosen because it roughly corresponds
to the fractional luminosity above which most of the discs are affected by the mass problem,
see Fig.~\ref{fig:mass&xm-fd_3.5_1.0mm_3.7_1.0km_2.8_200km_slide}.

The timescale $T_\mathrm{p}$, on which the progenitor gets disrupted in the disc,
is given by Eq.~(16) of \citet{wyatt-2008}.
The timescale $T_\mathrm{cl}$ on which the debris cloud decays with time
(considering only mutual collisions rather than with the background disc
as we require $f_\mathrm{cl} > f_\mathrm{bkg}$) can also be computed from
Eq.~(16) of \citet{wyatt-2008}.
Thus the ratio of the two timescales is
\be
{T_\mathrm{cl} \over T_\mathrm{p}}
=
{M_\mathrm{disc} \over \alpha M_\mathrm{p}}
\left( s_\mathrm{big} \over s_\mathrm{p} \right)
\left( s_\mathrm{p} \over s_\mathrm{max} \right)^{0.5}
=
\left( f_\mathrm{bkg} \over f_\mathrm{cl} \right)
\left( s_\mathrm{big} \over s_\mathrm{p} \right)^{0.5} .
\label{eq:timescale ratio}
\ee

The fractional luminosity of the collisional debris evolves as
$f_\mathrm{cl} / (1 + t / T_\mathrm{cl})$ 
\citep[see Eq.~17 of][]{wyatt-2008},
and so the time it spends
above $f_\mathrm{lim} = 10^{-4}$ is
$T_\mathrm{cl} (f_\mathrm{cl} / f_\mathrm{lim} - 1)$.
Thus the fraction of time for which the aftermath of a giant 
collision will have a fractional luminosity larger than $10^{-4}$,
is given by
$(T_\mathrm{cl}/T_\mathrm{p}) (f_\mathrm{cl} / f_\mathrm{lim} - 1)$,
and $T_\mathrm{cl}/T_\mathrm{p}$ from Eq.~(\ref{eq:timescale ratio})
is a good estimate of the fraction of time collisional debris from a single disruption event
should remain detectable.
It is the last equality in Eq.~(\ref{eq:timescale ratio})
 that leads to the conclusion that collisional debris that is
detectable above the background disc must be rare,
since by definition $f_\mathrm{bkg} / f_\mathrm{cl} < 1$
and so the only way of not making this rare is to have a large $s_\mathrm{big}$,
but Eq.~(\ref{eq:fcl})
shows that a large $s_\mathrm{big}$
would result in small $f_\mathrm{cl}$
and Eq.~(\ref{eq:frac lum ratio})
that this would also make it harder to achieve $f_\mathrm{cl} / f_\mathrm{bkg}>1$.

This fraction of time over which a disc has $f_\mathrm{cl} > 10^{-4}$
is plotted
in Fig.~\ref{fig:giant_impact} bottom.
A disc enhanced by impact-generated debris 
can only be sufficiently bright
for a non-negligible fraction of its age
if the background disc is massive (otherwise giant impacts are too rare and/or sufficiently
large progenitors are absent),
the disrupted progenitor is large (otherwise the amount of the injected debris is too low),
\emph{and} the largest fragment is large enough (otherwise the debris cloud decays too fast).
However, even in the most favourable case of those considered
(massive disc with $M_\mathrm{disc} \sim 100 M_\oplus$,
an Earth-sized progenitor with $s_0 \sim 10^4\km$,
boulder-sized largest debris fragment with $s_\mathrm{big} \sim 1\cm$),
the fraction of time when the disc is bright does not exceed
$\sim 10^{-6}$.
This fraction of time is far too small to explain all
those discs that require an unrealistically high mass, since 
about 20 percent of all stars have detectable debris discs
\citep{eiroa-et-al-2013,thureau-et-al-2014,montesinos-et-al-2016,sibthorpe-et-al-2018},
and about one-fourth or one-third of them are affected by the mass problem.

\subsection{``Planetesimals born small''?}
\label{ss:born small}

For planetesimals with a flat primordial distribution, the discussion in Sect.~\ref{ss:recent ignition}
and the collisional lifetimes plotted in
Fig.~\ref{fig:Tcoll} (with dashed lines) make it clear
that bodies larger than a kilometre in size are not part of the collisional 
cascade even in old systems, making only a minor contribution to the observed dust production 
through cratering collisions with much smaller projectiles.
As a result, setting the maximum size in collisional models to $\sim 1\km$
is sufficient to reproduce the observed discs.
In other words, large bodies are not required at all by collisional models.
However, since the disc mass is proportional to $s_\mathrm{max}^{4-q}$,
and $q = q_\mathrm{big} = 2.8$ at $s \ga 1\km$,
replacing $s_\mathrm{max} = 200\km$ with $s_\mathrm{max} = 1\km$
would reduce the total disc mass by a factor of $600$!

To see how this applies to the discs in our sample, we note that time-dependent size
$s_\mathrm{km}$ given by Eq.~(\ref{eq:skm}) serves as a reasonable estimate for the size of the
bodies that feed the cascade in these discs, given their age.
A caveat is that a derivation of $s_\mathrm{km}$
in Sect.~\ref{ss:recent ignition} assumed the flat primordial distribution. Yet its presence does not
alter collisional lifetimes of km-sized objects markedly, so that a calculation neglecting 
the bodies larger than kilometres would not change $s_\mathrm{km}$ more than by a factor of a few.
This size, which is plotted in
Fig.~\ref{fig:skm_and_Mdisc} (top) and listed in Table~\ref{tab:sample}, ranges from
$\approx 0.1\km$ to $\approx 7\km$.
Assuming now that bodies larger than $s_\mathrm{km}$ were absent, the discs would have a mass
\be
 M_\mathrm{disc}^\mathrm{min}
 = 
 M_\mathrm{d}
   \;
   {4 - q\phantom{_big} \over 4 - q_\mathrm{med}}
 \left(s_\mathrm{km} \over s_\mathrm{mm} \right)^{4-q_\mathrm{med}} .
\label{eq:min disc mass}
\ee
This mass is given in the last column of Table~\ref{tab:sample} and ranges from $0.2 M_\oplus$
(for the $\varepsilon$~Eri disc) to $86 M_\oplus$ (for the HD~131488 disc).
These values are within an order of magnitude consistent with ``minimum disc mass'' estimates
in Sect.~\ref{ss:min mass} obtained
with the mass loss argument and are a way below the ``maximum disc masses'' discussed in
Sect.~\ref{ss:max mass}.
In summary, a hypothesis that 
large bodies for some reason do not form on the periphery of planetary systems would 
provide the easiest solution to the debris disc mass problem.

This conclusion is also robust with respect to a possible uncertainty in the assumed
slope $q_\mathrm{med}$.
This is because $s_\mathrm{km}$ in Eq.~(\ref{eq:min disc mass}) also depends on 
$q_\mathrm{med}$ in such a way that $M_\mathrm{disc}^\mathrm{min}$ is independent of $q_\mathrm{med}$.
Indeed, for any size distribution with a single slope $3 < q_\mathrm{med} < 4$ that is in steady state,
the collisional lifetime of an object of size $s$ is
$T_\mathrm{coll}(s) = T_\mathrm{coll}(s_\mathrm{mm})(s/s_\mathrm{mm})^{4-q_\mathrm{med}}$
\citep[see, e.g., Eq. 36 in][]{wyatt-et-al-2011}.
Since $s_\mathrm{km}$ should satisfy the equation
$T_\mathrm{coll}(s_\mathrm{km}) = T_\mathrm{age}$,
we find
$(s_\mathrm{km}/s_\mathrm{mm}) = [T_\mathrm{age}/T_\mathrm{coll}(s_\mathrm{mm})]^{1/(4-q_\mathrm{med})}$.
Substituting this into Eq.~(\ref{eq:min disc mass}) shows that 
$M_\mathrm{disc}^\mathrm{min}$ is independent of $q_\mathrm{med}$ in the single-slope approximation.
In fact, the conclusion that $M_\mathrm{disc}^\mathrm{min}$ is independent of $q_\mathrm{med}$ is more general and
only relies on the assumption of steady state.
This is because in steady state the mass loss rate from the top end of the cascade
($M_\mathrm{disc}^\mathrm{min} / T_\mathrm{age}$)
is equal to that at the small size end
($M_\mathrm{d} / T_\mathrm{coll}(s_\mathrm{mm})$).
The derived $M_\mathrm{disc}^\mathrm{min}$
is thus completely independent of the size distribution
above $s_\mathrm{mm}$.

\begin{figure}
\includegraphics[width=1.05\columnwidth]{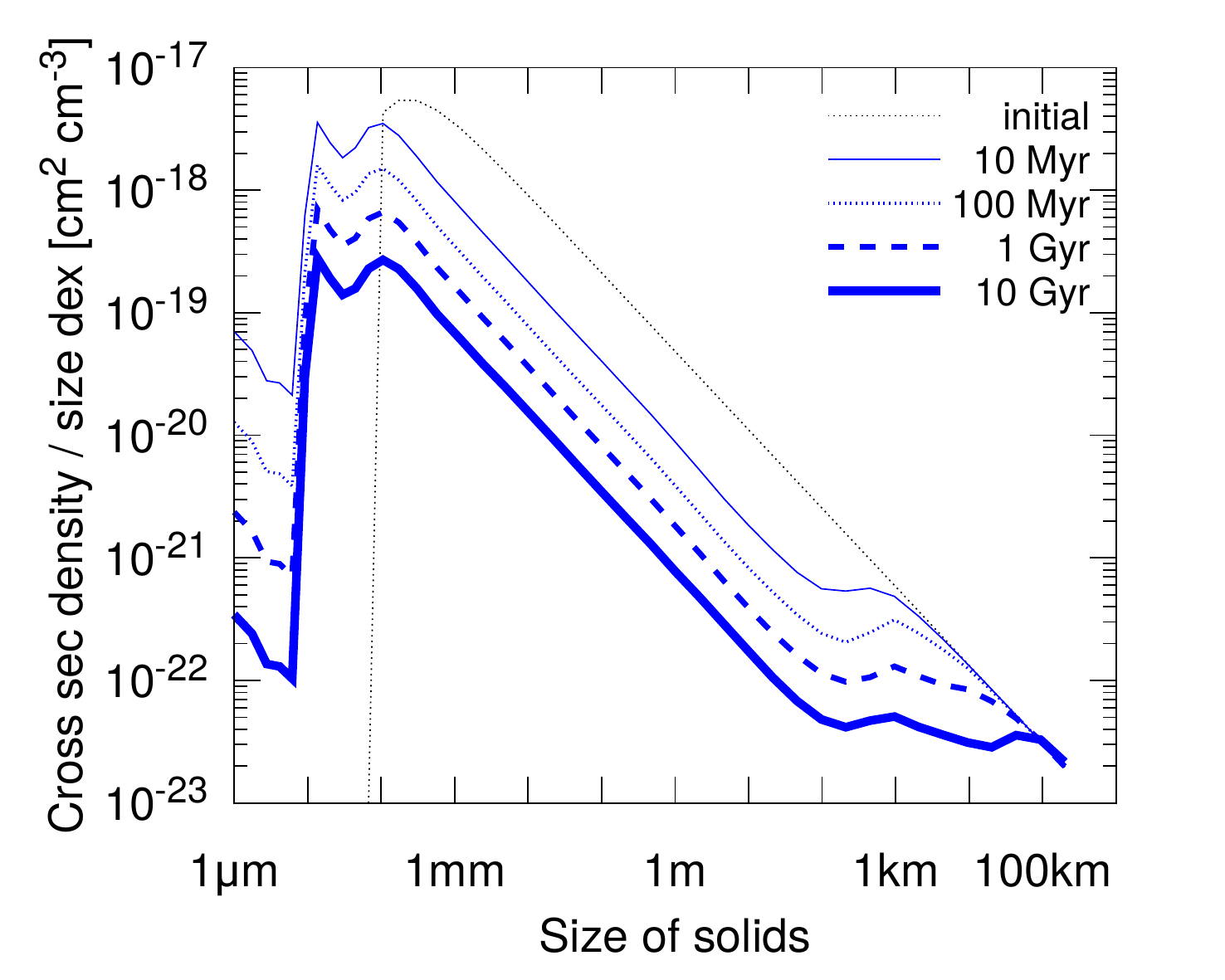}\\
\includegraphics[width=1.05\columnwidth]{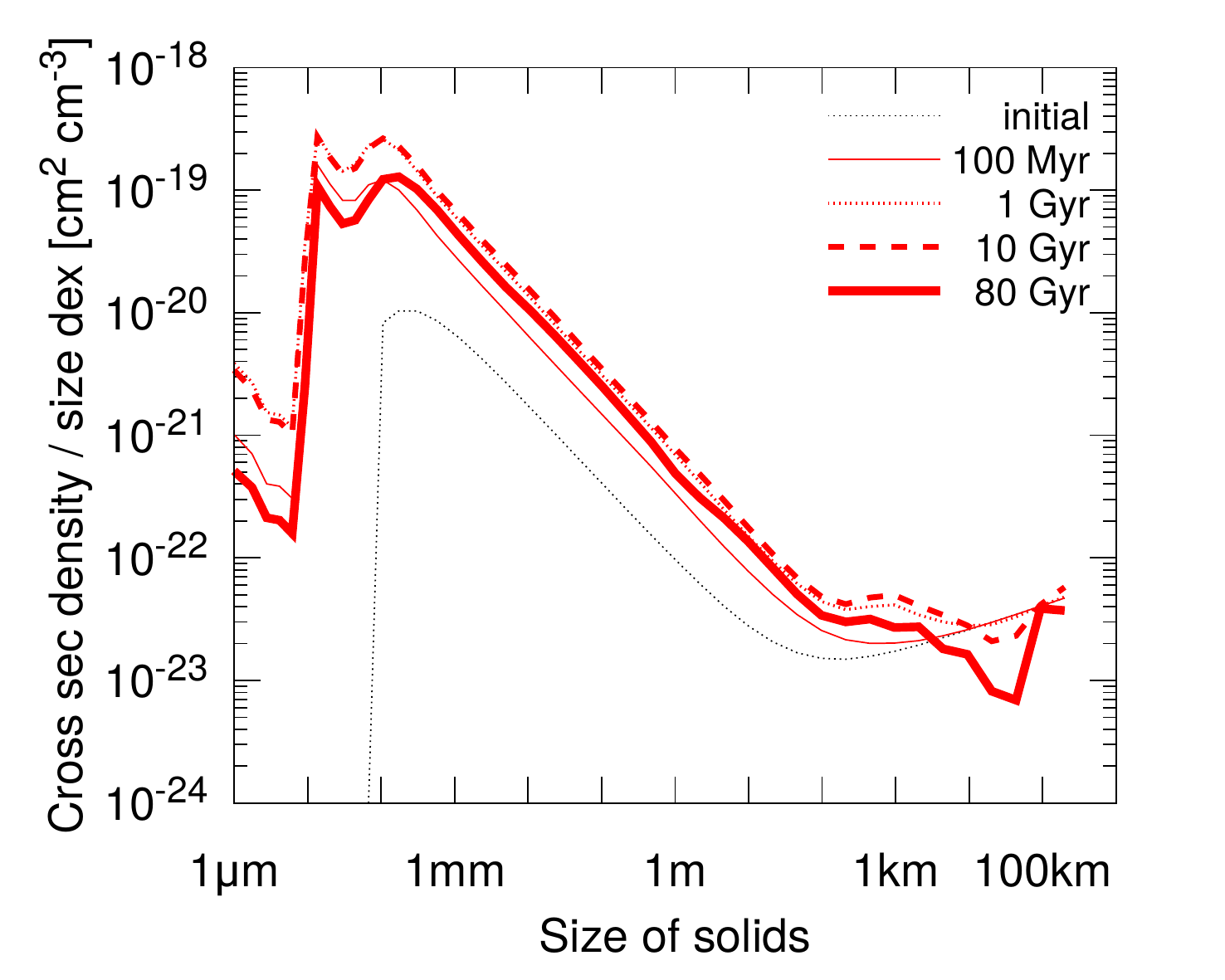}
    \caption{Long-term evolution of the size distribution in two fiducial debris discs
    of $100 M_\oplus$ (Appendix~\ref{ss:ACE sims}),
    assuming different size distribution slopes for a primordial population of planetesimals:
    $q_\mathrm{big} = 3.7$ (run~A, top panel) and $q_\mathrm{big} = 2.8$ (run~B, bottom).
    Black dotted line: initial distributions.
    Coloured lines of increasing thickness: distributions at later times, as indicated in the legends.
    }
    \label{fig:longterm}
\end{figure}

\looseness=-1
The above discussion makes it obvious that, if the bodies larger than $s_\mathrm{km}$ were absent,
the mass problem would not arise.
There is also an argument against the presence of a primordial population
of planetesimals with a relatively flat size distribution
from the fact that such a primordial distribution results in dust mass that increases with age.
Given the three-phase size distribution discussed in Sect.~\ref{ss:formation},
with $s_\mathrm{km}$ increasing with age as in Sect.~\ref{ss:recent ignition},
this increase of dust mass with age occurs whenever $q_\mathrm{big}<q_\mathrm{med}$,
and continues until the largest planetesimals reach collisional equilibrium.
This behaviour is reproduced in the numerical simulations of the size distribution
(run~B in Appendix~\ref{ss:ACE sims}),
which show 
that while a three-phase distribution is an approximation,
there is still an increase in dust mass until $\sim 10\Gyr$
(Fig.~\ref{fig:longterm} bottom).
This contradicts observations,
as there is ample evidence for a long-term decay in the dust mass in debris discs
\citep[e.g.,][]{zuckerman-becklin-1993,habing-et-al-1999,habing-et-al-2001,spangler-et-al-2001,%
greaves-wyatt-2003,rieke-et-al-2005,moor-et-al-2006,eiroa-et-al-2013,chen-et-al-2014,%
moor-et-al-2014,holland-et-al-2017,sibthorpe-et-al-2018}.

We note that collisional equilibrum at large sizes is reached, and so the dust mass starts 
to decay, on much shorter timescales in smaller discs. For instance, \citet{nesvorny-vokrouhlicky-2019} did a 
collisional simulation for a primordial Kuiper belt of $40 M_\oplus$ at around $\sim 25\AU$
with an initial size distribution from streaming instability simulations, similar to our run~B in
Appendix~\ref{ss:ACE sims}. They found that collisional equilibrium is achieved by bodies of
up to $50\km$ in radius in just $\sim 50\Myr$. Equation~(\ref{eq:Tage}) suggests that the collisional 
timescale scales as $M_\mathrm{disc}^{-1} M_\star^{-1.33} R^{4.33}$. Thus, if our run~B gave $\sim 10\Gyr$
for $200\km$-sized planetesimals in a $100 M_\oplus$ disc with a radius of $100\AU$ around a $2.16 
M_\odot$ star, then same-sized bodies in the Kuiper belt would reach equilibrium at $\sim 170 \Myr$.
For $50\km$-sized bodies, the timescale would be even shorter, consistent with their result.
What this means is that large planetesimals may still be required
(i.e., $s_\mathrm{km} \gg 1\km$) to replenish the dust in discs with small radii found around old stars.
Moreover, for small discs it may be concluded for old stars that the level of dust observed
is incompatible with steady state erosion regardless of the size of their planetesimals
\citep[e.g.,][]{wyatt-et-al-2007}.

To make the dust mass decrease in time for all discs regardless of their radius, we have to admit
that the primordial population has
a sufficiently steep size distribution
(as in run~A of Appendix~\ref{ss:ACE sims}, see Fig.~\ref{fig:longterm} top).
To find out what ``sufficiently steep'' means, we note that the slope $q_\mathrm{med}$
flattens from $3.7$ for objects smaller than $\sim 0.1\km$ that reside in the strength regime
(as predicted by Eq.~\ref{eq:q(gamma,p)} with $\gamma = -0.37$ and $p=0$)
to $3.0$ for larger objects that are in the gravity regime (where $\gamma = 1.38$).
It is the latter value that determines how steep
the primordial size distribution of planetesimals has to be in order not to cause
the long-term increase in the amount of dust that contradicts the observations.
In other words, for the debris disc brightness to decrease with time as observed,
this requires the initial size distribution of mass-dominating large planetesimals
to be steeper than $q_\mathrm{big}=3.0$.

\looseness=-1
To solve the disc mass problem, we also need the size distribution
to be such that the ratio of the total mass to the mass in km-sized objects
(i.e., the largest objects in the primordial population that are small enough
to experience catastrophic collisions on timescales shorter than the system ages)
is not too large.
This can be achieved either by a very steep distribution
($q_\mathrm{big} > 4$ to put the mass in the smallest objects)
or for the largest planetesimals to be smaller than assumed.
While the idea that large planetesimals are nearly or completely absent
in bright debris discs would help
solve the mass problem and would remove any possible tension with the observed statistics
of different-aged discs, it comes with several caveats.
One is that rapid and efficient
formation of planetesimals as big as hundreds of kilometres is
robustly predicted by certain models of planetesimal formation. For instance, 
\citet{schaefer-et-al-2017} predict the top-end of the initial mass distribution
to be exponentially tapered at masses $\sim 10^{-5} G^{-1} H_\mathrm{g}^3 P^{-2}$
(with $G$ being the gravitational constant and $H_\mathrm{g}$ the scale height of gas in a 
protoplanetary disc). For a broad range of plausible parameters, this corresponds
to sizes on the order of hundreds of kilometres.
Nevertheless, many aspects of planetesimal formation are not yet fully understood.
It appears possible that state-of-the-art models might still miss some essential pieces of
physics or that some of the underlying assumptions fail in reality.

Another concern about the suggested scarcity of $>1\km$ bodies in debris discs
might be the fact that our own Kuiper belt is populated by numerous
transneptunian objects $\sim 100\km$ in size.
While Kuiper-belt objects (KBOs) in the dynamically hot populations (hot classicals,
scattered disc objects) must have formed closer in $<30\AU$ and were implanted in
the present-day Kuiper belt by the migration of Neptune,
there is ample evidence that the cold classical KBOs accreted in-situ at their current
location between $42$--$48\AU$ \citep[see][and references therein]{morbidelli-nesvorny-2020}.
This conclusion is strongly backed up, for instance, by a high fraction of equal-sized
binaries in this population \citep[e.g.,][]{nesvorny-vokrouhlicky-2019}.
Thus we do know that planetesimals with $\sim 100\km$ in size formed successfully
at the periphery of our Solar system. Their inferred size distribution between a
few km and $\sim 50$--$100\km$ is flat ($q_\mathrm{big} \sim 3$),
in good agreement with planetesimal formation
model predictions. Above $\sim 50$--$100\km$, the distribution steepens drastically
to $q_\mathrm{big} \ga 6$ \citep{bernstein-et-al-2004}, so that the largest cold classicals such as
(119951) 2002 KX14 with $s \sim 200\km$ \citep{vilenius-et-al-2012} make a
negligible contribution to the total mass. Nevertheless, these facts are only valid for
our own Solar system (which is not a system with a disc mass problem), and we do not know whether it is representative of other planetary systems.
Given the bewildering diversity of extrasolar planets discovered so far,
it is quite possible that circumstances and outcomes of planetesimal formation also vary
significantly from one system to another.

\section{Discussion}
\label{s:discussion}

\subsection{A possible ``big picture''}

The above analysis points to a conclusion that large planetesimals should be absent in systems
with bright debris discs.
This implies that planetesimal formation processes yield different-sized planetesimals in some systems.
The reasons for that are not known, but they could be related to the conditions in the protoplanetary discs.
It is also possible that the efficiency of planetesimal formation, i.e.,
the fraction of available mass that is turned into planetesimals of some size,
could be different for the formation of smaller versus larger planetesimals.

A visual inspection of plots such as Fig.~\ref{fig:mass&xm-fd_3.5_1.0mm_3.7_1.0km_2.8_200km_slide}
suggests that the total mass of debris discs, inferred under an assumption of $s_\mathrm{max}$ being
the same for all discs, shows a monotonic rise with fractional luminosity. One hypothesis
could be that the true total mass in all the discs is comparable but, for some reasons, the largest
planetesimals in brighter discs are smaller.
Indeed, Eq.~(\ref{eq:disc mass}) tells us the inferred total mass is proportional to 
$M_\mathrm{d} s_\mathrm{max}^{4-q_\mathrm{big}}$, where
$M_\mathrm{d} \propto f_\mathrm{d}$.
Were $s_\mathrm{max} \propto f_\mathrm{d}^{-1/(4-q_\mathrm{big})}
\propto f_\mathrm{d}^{-0.8}$, the total mass of all discs would be similar.
To give a numerical example, all bright discs with fractional luminosities in the range from
$10^{-4}$ to $10^{-2}$  could have approximately the same mass
$M_\mathrm{disc} \sim 100 M_\oplus$, if the radius of the biggest planetesimals
varied from $\sim 200\km$ at $f_\mathrm{d} = 10^{-4}$
to $\sim 4\km$ at $f_\mathrm{d} = 10^{-2}$.

In other words, it can be that all protoplanetary discs leave behind a comparable mass
in small bodies by the time of gas dispersal, yet the ability to form the largest objects
varies from one system to another.
In that case, the brightest debris discs would form in those systems where the planetesimals
have smaller, kilometre sizes.
And conversely, fainter debris discs would emerge in
those systems where planetesimals hundreds of kilometres in radius succeeded to form.

Our present-day Kuiper belt would be exempt from this trend,
as its total mass is much lower than that of the (currently detectable) extrasolar discs.
The likely reason for that is early depletion of the primordial Kuiper belt by a dynamical instability
of giant planets \citep[e.g.,][]{morbidelli-nesvorny-2020}.
Without that event, the $\sim 30 M_\oplus$ debris disc of the Solar system would have been
comparable in brightness with the brightest known extrasolar discs in terms
of its $24$\,$\mu$m and $70$\,$\mu$m emission \citep[see Fig.~12 in][]{booth-et-al-2009}.
This may appear incompatible with the above argument in which a bright primordial Kuiper belt
might be taken to infer that it was born with small planetesimals,
whereas these models include bodies up to $500\km$ in radius.
However, the primordial Kuiper belt's infrared emission is not inferred to be bright
because its mass is concentrated in small planetesimals,
rather because of its small radius ($\sim 20 \ldots 30\AU$) and so hot temperature
compared with extrasolar discs which are more typically $\sim 100\AU$.
This underscores the importance of disc radius in the analysis,
since being bright in the infrared does not necessarily indicate a mass problem if the emission is hot.
Sub-mm dust mass is a better indicator of a possible disc mass problem,
since by that metric the primordial Kuiper belt is one to two orders of magnitude lower
than that of the brightest known debris discs
\citep[see Fig.~12 in][for their ``comet-like'' dust composition]{booth-et-al-2009}.

It is also interesting that the discs that appear to be the most problematic and so
need the ``planetesimals born small'' hypothesis,
are those with the largest radii
(see Fig.~\ref{fig:mass vs various}a).
This suggests that the biggest planetesimals may be smaller at larger
distances. This possibility might be supported by recent simulations of \citet{klahr-schreiber-2020},
which showed that the expected typical size of planetesimals formed via the SI plunges down
outside $\sim 50\AU$ from the star (see their Fig.~10).
Thus it is possible that more compact discs, such as the Kuiper belt, succeeded to form larger planetesimals,
whereas many of the well known extrasolar discs, being much larger, have a size distribution limited
to smaller ones.

Of course, there is no requirement for all stars to have discs with the same mass.
Just as the Kuiper belt is lower in mass because of some level of depletion,
so too could the $f_\mathrm{d} = 10^{-4}$ discs simply be lower in mass,
either because they formed less massive,
or because they suffered some depletion.
As a numerical example, an $f_\mathrm{d} = 10^{-4}$ disc could have a total mass as low as 
$M_\mathrm{disc} \sim 1 M_\oplus$, if the biggest planetesimals were as small as
$\sim 4\km$ in radius.

\subsection{Implications}

\looseness=-1
One immediate implication of having fewer large planetesimals in debris discs would be
that these discs are not self-stirred \citep{kenyon-bromley-2008,kennedy-wyatt-2010,krivov-booth-2018}.
For instance, the presence of $\ga 400\km$-sized embedded bodies is required to self-stir
the AU~Mic disc \citep{daley-et-al-2019}, and at least $\sim 100\km$-sized planetesimals
are necessary to reproduce the stirring level of the dynamically cold population
recently discovered in the $\beta$~Pic disc \citep{matra-et-al-2019b}.
However, there are possibilities other than self-stirring to explain the inferred excitation
level of planetesimals in debris discs.
Alternative mechanisms include stirring by planets near the inner edges
of the discs \citep{quillen-2006a} or elsewhere in the disc cavities
\citep{mustill-wyatt-2009} or ```primordial'' excitation of planetesimals forming in
self-gravitating protoplanetary discs \citep{walmswell-et-al-2013}.

A lack of big planetesimals would also affect planetesimal-driven migration of planets
in the systems, which is known to be a viable mechanism to explain structure
in our Kuiper belt \citep{malhotra-1995} and extrasolar debris discs \citep{wyatt-2003}.
Migration, which otherwise has a stochastic component \citep{zhou-et-al-2002}, would become smoother
in the absence of large bodies \citep{chiang-et-al-2003}.
This would make resonant trapping of planetesimals easier, perhaps helping to explain some
of the observed features in the discs such as the $\beta$~Pictoris clump \citep{dent-et-al-2014}.

Should the systems with bright debris discs indeed be dominated by small (km-sized) planetesimals,
this would necessitate revisions to the existing planetesimal formation models.
Perhaps this could also help constrain conditions in the PPDs and how they vary from one system to another.
Strictly speaking, our conclusion only applies to the outer region (outside 10s to $\AU$)
that is currently occupied by debris-producing residual planetesimals. It is not clear
if the planetesimal formation that happens at large distances is similar
to what is going on closer in.
If so, this would have profound implications
for planet formation theories (e.g., growth to embryos and full-size planets).
Specifically, this would pose a challenge of how to build inner planets~---
which we know to exist in many systems~--
starting from planetesimals of smaller sizes
\citep[see a discussion in][and references therein]{voelkel-et-al-2020}.

\subsection{Potential tests}

\looseness=-1
Are there any ways to constrain either the sizes of the largest planetesimals,
the total masses of debris discs, or both?
In principle information on the largest planetesimals
is encoded in the statistics of debris disc detections as a function of age.
Population models which follow the erosion of planetesimal belts as they deplete
through collisions have been able to reproduce these statistics
\citep[e.g.,][]{wyatt-et-al-2007,loehne-et-al-2007,gaspar-et-al-2013}.
However, they have yet to set strong constraints on the largest planetesimal size or disc mass,
since the collisional depletion timescale in such models is degenerate with the level
of stirring and planetesimal strength for example.
The distribution of planetesimal belt radii is also very uncertain when inferring
this from the emission spectrum, which is another important parameter in determining
collisional lifetimes. High resolution imaging is improving constraints on stirring levels
\citep[e.g.,][]{daley-et-al-2019,matra-et-al-2019b} and also on planetesimal belt radii
which may correlate with stellar luminosity \citep{matra-et-al-2018}.
This means that we can hope that constraints on parameters such as the size
of the largest planetesimals will soon be possible.
For example, it can be expected that discs in which the planetesimals
are born small would appear bright at young ages (10s of Myr) then rapidly decay
(e.g., within a few 100~Myr), whereas those in which planetesimals are born large
would start out fainter, but maintain that brightness at a relatively constant level
over several Gyr.
Thus it may be possible to find signatures of the largest planetesimal size
in comparative studies of populations in the 10Myr--Gyr range.

As far as measuring disc mass is concerned, one possibility is to use the observed disc
structure as an indicator of disc mass.
For example, if the disc is massive enough for its
self-gravity to become important, this could affect the structure of the disc (Sefilian \& Wyatt, in prep.).
Besides, planet-disc interaction should result in different dynamical patterns, depending on whether
the disc mass is lower or comparable to that of the planet \citep{pearce-wyatt-2014,pearce-wyatt-2015}.
For instance, in the latter case a single planet in eccentric orbit could explain a double-ringed structure
of a debris disc exterior to the planet's orbit, such as the one observed in
HD 107146 \citep{marino-et-al-2018} and HD 92945 \citep{marino-et-al-2019}.
In the former case, the same structure could rather be attributed to a planet in a nearly-circular orbit between
the two rings and clearing its orbit from debris.
Thus finding planets in one of those systems would help indirectly ``measure'' the disc mass,
depending on the orbit and mass of the planet discovered.
Finally, a disc's mass would also affect how far and how fast (Neptune-mass) planets migrate through
planetesimal scattering, which may have observable consequences
\citep[see, e.g., Sect.~4.3 in][for a detailed discussion]{marino-et-al-2018}.

\section{Conclusions}
\label{s:conclusions}

In this paper, we consider the total mass of debris discs and how it is related
to the overall mass budget of solid material in planetary systems.
Since the disc mass is dominated by directly unobservable planetesimals,
it has to be estimated from the mass of debris dust produced in their collisions.
That dust mass is inferred from (sub)mm observations, assuming a certain opacity.
The dust mass is then extrapolated to the mass of planetesimals up to a certain size,
using theoretical models of a dust-producing collisional cascade.

The total disc mass found in this way must satisfy certain constraints that we define as follows.
The ``minimum debris disc mass'', which is needed to collisionally sustain
the debris dust at the observed level over the system's age, is estimated to be
$\sim 10 M_\oplus$.
The ``maximum debris disc mass'' is determined by the total mass of condensible compounds
that are available in protoplanetary discs out of which debris discs and planets formed.
This should be of the order of
 $\sim 100$...$1000 M_\oplus$.

Estimating the dust mass for a sample of 20 bright debris discs
with reliable ALMA (sub)mm fluxes,
and extrapolating up to the $200\km$-sized planetesimals,
we find the total disc mass to lie in the  $\sim 10^3$...$10^4 M_\oplus$ range
for the majority of discs in the sample, which clearly exceeds the maximum possible mass.
We refer to this as a ``debris disc mass problem.''

In an attempt to resolve the controversy, we re-analyse possible uncertainties
as well as the various assumptions made in estimating the disc mass. For instance, the
dust mass is uncertain by a factor of several, because so is the assumed dust opacity.
It is also possible that collisional damping
and damping by secondary gas steepen the size distribution somewhat, although we find these
processes to be inefficient.
Next, if the bodies are weaker than assumed (which increases the size of the bodies
that start to be depleted by the age of the systems),
the estimated total disc mass will reduce by a factor of several.
If the density of larger bodies is lower than that of smaller ones,
this will also result in a several times smaller disc mass estimate.
None of the factors like these, taken alone, would solve the mass
problem. However, it cannot be ruled out that several of them in combination would mitigate
the tension between the estimated and maximum possible debris disc masses, although probably
not providing an ultimate solution to the problem.

We argue that the easiest solution to the mass problem would be to admit that the size of the
largest planetesimals in bright discs is of the order of kilometres, and thus
smaller than might be expected given the size distribution of planetesimals in the Solar system
and current planetesimal formation models.
It is possible that the largest planetesimals formed farther out from the star have smaller sizes,
so that compact discs such as our Kuiper belt are dominated by large planetesimals, while
larger discs have a size distribution limited to small planetesimals.
The size of the biggest bodies may also vary strongly from one system to another,
such that the systems where planetesimal accretion far from the star stalled at smaller sizes
formed the brightest debris discs.
This ``planetesimals born small'' scenario would set important constraints on the planetesimal formation
processes from the debris disc perspective. It would also have a number of implications
for planetesimal-driven migration and growth of planetary cores and planets.

\vspace*{-3mm}
\section*{Data availability}

The data underlying this article will be shared on reasonable request to the corresponding author.

\section*{Acknowledgements}

We are grateful to Luca Matr{\`a} for providing us with (sub)mm fluxes
from his analysis of the sample of ALMA- and SMA-resolved debris discs
used in this paper, Torsten L\"ohne for sharing with us the optical data
for a selection of materials,
and Hiroshi Kobayashi for useful discussions on collisional damping.
An insightful and speedy review report by Alessandro Morbidelli is very much appreciated.
AVK acknowledges support from the {\em Deutsche Forschungsgemeinschaft} (DFG) through
grant Kr~2164/13-2. 



\newcommand{\AAp}      {Astron. Astrophys.}
\newcommand{\AApR}     {Astron. Astrophys. Rev.}
\newcommand{\AApS}    {AApS}
\newcommand{\AApSS}    {AApSS}
\newcommand{\AApT}     {Astron. Astrophys. Trans.}
\newcommand{\AdvSR}    {Adv. Space Res.}
\newcommand{\AJ}       {Astron. J.}
\newcommand{\AN}       {Astron. Nachr.}
\newcommand{\AO}       {App. Optics}
\newcommand{\ApJ}      {Astrophys. J.}
\newcommand{\ApJL}      {Astrophys. J. Lett.}
\newcommand{\ApJS}     {Astrophys. J. Suppl.}
\newcommand{\ApSS}     {Astrophys. Space Sci.}
\newcommand{\ARAA}     {Ann. Rev. Astron. Astrophys.}
\newcommand{\ARevEPS}  {Ann. Rev. Earth Planet. Sci.}
\newcommand{\BAAS}     {BAAS}
\newcommand{\CelMech}  {Celest. Mech. Dynam. Astron.}
\newcommand{\EMP}      {Earth, Moon and Planets}
\newcommand{\EPS}      {Earth, Planets and Space}
\newcommand{\GRL}      {Geophys. Res. Lett.}
\newcommand{\JGR}      {J. Geophys. Res.}
\newcommand{\JOSAA}    {J. Opt. Soc. Am. A}
\newcommand{\MemSAI}   {Mem. Societa Astronomica Italiana}
\newcommand{\MNRAS}    {MNRAS}
\newcommand{\PASJ}     {PASJ}
\newcommand{\PASP}     {PASP}
\newcommand{\PSS}      {Planet. Space Sci.}
\newcommand{\QJRAS}    {Quart. J. Roy. Astron. Soc.}
\newcommand{\RAA}      {Research in Astron. Astrophys.}
\newcommand{\SolPhys}  {Sol. Phys.}
\newcommand{\SolSysRes}{Sol. Sys. Res.}
\newcommand{\SSR}      {Space Sci. Rev.}

\input ms.bbl




\appendix
\section{Collisional simulations}
\label{ss:ACE sims}

In this Appendix, we describe two simulations with the ACE code
\citep[e.g.,][]{krivov-et-al-2013} which we refer to throughout the paper:

\begin{itemize}

\item
{\bf Run~A}.
The setup included a $10\AU$ wide ring of planetesimals around a distance of $100\AU$.
As a central star, we chose an A-type star of $2.16$ solar masses with a luminosity of
$27.7$ times the solar one.
For planetesimals smaller than $1\km$, we took the initial size distribution with
a slope $q_\mathrm{med} = 3.7$.
The largest planetesimals were assumed to have $200\km$ in radius,
and the initial size distribution slope of the
primordial population of objects with radii from $1\km$ to $200\km$
was set to $q_\mathrm{big} = 3.7$, i.e., made the same as for sub-km planetesimals.
The total disc mass was set to $100M_\oplus$.
The planetesimals were assumed to have a distribution of eccentricities
between 0 and 0.1 and inclinations from 0 to 3 degrees.
As material, we took a mixture of astrosilicate \citep{draine-2003} and water ice
\citep{li-greenberg-1998} in equal volume fractions with a bulk density of $\rho = 2.35 \g\cm^{-3}$.
A number of other standard assumptions were also made
\citep[see][for a complete list of parameters assumed]{krivov-et-al-2018}.

Results of run~A are shown in
Fig.~\ref{fig:dohnanyi} (a typical simulated size distribution),
Fig.~\ref{fig:Tcoll} (with solid lines; collisional lifetimes),
and
Fig.~\ref{fig:longterm} top (long-term evolution of the size distribution).

\item
{\bf Run~B}.
All the parameters were taken the same as in run~A, except that
the assumed initial size distribution slope
of the primordial population of objects with radii from $1\km$ to $200\km$
was set to $q_\mathrm{big} = 2.8$,
as predicted by pebble concentration models of planetesimal formation.

Results of run~B are shown in
Fig.~\ref{fig:pebble-piles} (a typical simulated size distribution),
Fig.~\ref{fig:Tcoll} (with dashed lines; collisional lifetimes),
and
Fig.~\ref{fig:longterm} bottom (long-term evolution of the size distribution).

\end{itemize}


\bsp	
\label{lastpage}
\end{document}